\documentclass[a4paper,10pt]{article}
\usepackage[dvipdfmx]{graphicx,color}
\usepackage{amssymb}
\usepackage{amsmath}
\usepackage{latexsym}
\usepackage{array}
\usepackage{bm}
\usepackage{ascmac}

\def\qed{\hfill\hbox{\rule[-2pt]{3pt}{6pt}}}

\newtheorem{thm}{Theorem}[section]

\newtheorem{prop}[thm]{Proposition}

\newtheorem{dfn}[thm]{Definition}

\newtheorem{proof}{Proof}

\makeatletter

\@addtoreset{equation}{section}
\makeatother

\title{Bilinear equations and $q$-discrete Painlev\'e equations satisfied by variables and coefficients in cluster algebras}
\author{Naoto Okubo\\Graduate School of Mathematical Sciences, the University of Tokyo, \\3-8-1 Komaba, Tokyo 153-8914, Japan}
\date{}

\begin{document}
\maketitle

\begin{abstract}
We construct cluster algebras the variables and coefficients of which satisfy the discrete mKdV equation, the discrete Toda equation and other integrable bilinear equations, several of which lead to $q$-discrete Painlev\'e equations.
These cluster algebras are obtained from quivers with an infinite number of vertices or with the mutation-period property.
We will also show that a suitable transformation of quivers corresponds to a reduction of the difference equation.
\end{abstract}

\section{Introduction}

In this article, we deal with cluster algebras, which were introduced by Fomin and Zelevinsky \cite{1,2}.
A cluster algebra is a commutative ring described by cluster variables and coefficients.
A generating set of the cluster algebra is defined by mutation, which
is a transformation of a seed consisting of a set of cluster variables, coefficients, and a quiver.
Cluster variables and coefficients obtained from a mutation of an initial seed satisfy some difference equations.
It is known that cluster variables can satisfy the discrete KdV equation \cite{4} and the Hirota-Miwa equation \cite{5}, when the initial seed includes suitable quivers with infinite vertices \cite{6}.
These quivers have the property that an infinite number of mutations gives a permutation of its vertices.
This property is called `mutation-period' and a quiver with this property is called a mutation-periodic quiver.
Several results concerning mutation-periodic quivers have been reported in \cite{3}.
All the mutation-periodic quivers in which a permutation of its vertices is achieved by a single mutation have already obtained.
The quiver which gives the discrete KdV equation is obtained from a transformation of the quiver of the Hirota-Miwa equation.
This transformation corresponds to a reduction from the Hirota-Miwa equation to the discrete KdV equation.
In this paper, we construct the cluster algebras whose variables and coefficients satisfy the discrete mKdV equation, the discrete Toda equation \cite{7}, and some $q$-discrete Painlev\'e equations \cite{8}.
We introduce the quiver which generalizes the one that corresponds to the discrete KdV equation and the Hirota-Miwa equation.
Quivers of $q$-Painlev\'e I,II equations and their higher order analogues have been obtained in \cite{6,9}.
We shall introduce the quivers for the $q$-Painlev\'e III,VI equations, which are mutation-periodic and are obtained from transformations of quivers for the discrete KdV equation and the discrete mKdV equation.

\section{Cluster algebras}

In this section, we briefly explain the notion of cluster algebra which we use in the following sections.
Let $\bm{x}=(x_1,x_2,\dots ,x_N), \bm{y}=(y_1,y_2,\dots ,y_N)$ be $N$-tuple variables.
Let $Q$ be a quiver with $N$ vertices.
Consider the quiver whose vertices correspond to the cluster variables.
We assume that the quiver does not have a loop ($i\longrightarrow i$) or a 2-cycle ($i\longrightarrow j\longrightarrow i$).
Each $x_i$ is called a cluster variable and each $y_i$ is called a coefficient.
The triple $(Q,\bm{x},\bm{y})$ is called a seed.
Let $\lambda_{i,j}$ be the number of arrows from $i$ to $j$ of the quiver $Q$.
We define $\lambda_{i,j}$ for all $1\leq i,j\leq N$ as $\lambda_{j,i}=-\lambda_{i,j}$.
As a quiver $Q$ does not have a loop, $\lambda_{i,i}=0$.
A mutation is a particular transformation of seeds.
\begin{dfn}\label{dfn:ca}
Let $\mu_k :(Q,\bm{x},\bm{y})\longmapsto(Q',\bm{x'},\bm{y'})\quad (k=1,2,\dots ,N)$ be the mutation at the vertex $k$ of the quiver $Q$, defined as follows.
\begin{itemize}
\item $Q'$ is a new quiver, obtained by three operations on the quiver $Q$.
\begin{enumerate}
\item For all $(i,j)$ such that $\lambda_{i,k}>0,\lambda_{k,j}>0$, we add $\lambda_{i,k}\lambda_{k,j}$ arrows from $i$ to $j$.
\item If 2-cycles appear by the operation 1, we remove all of them.
\item We reverse the direction of all directed arrows which have edges at the vertex $k$.
\end{enumerate}
\item New coefficients $\bm{y'}=(y_1',y_2',\dots ,y_N')$ are defined from $Q$ and $\bm{y}$ as:
\begin{equation}\label{coef}
\begin{aligned}
y_k'&=y_k^{-1},&\\
y_i'&=y_i\left(y_k^{-1}+1\right)^{-\lambda_{k,i}}\quad &(\lambda_{k,i}>0),\\
y_i'&=y_i(y_k+1)^{\lambda_{i,k}}\quad &(\lambda_{i,k}>0),\\
y_i'&=y_i\quad &(\lambda_{k,i}=0).
\end{aligned}
\end{equation}
\item New cluster variables $\bm{x'}=(x_1',x_2',\dots ,x_N')$ are defined from $Q$, $\bm{y}$ and $\bm{x}$ as:
\begin{equation}\label{clus2}
\begin{aligned}
x_k'&=\frac{1}{(y_k+1)x_k}\left( \prod_{\lambda_{k,j}>0}x_j^{\lambda_{k,j}}
+y_k\prod_{\lambda_{j,k}>0}x_j^{\lambda_{j,k}}\right),\\
x_i'&=x_i\quad (i\neq k).
\end{aligned}
\end{equation}
\end{itemize}
\end{dfn}
A mutation $\mu_k$ denotes the mutation at $k$ or $x_k$.
For any seed $(Q,\bm{x},\bm{y})$, it holds that $\mu_k^2(Q,\bm{x},\bm{y})=(Q,\bm{x},\bm{y})$.
For any seed $(Q,\bm{x},\bm{y})$ and $(i,j)$ such that $\lambda_{i,j}=0$, it holds that $\mu_i\mu_j(Q,\bm{x},\bm{y})=\mu_j\mu_i(Q,\bm{x},\bm{y})$.
\begin{dfn}
Let us fix a seed $(Q,\bm{x},\bm{y})$.
This seed is called an initial seed.
Let $\mathcal{A}(Q,\bm{x},\bm{y})$ be a cluster algebra (with coefficients) defined as
\begin{equation}
\mathcal{A}(Q,\bm{x},\bm{y})=\mathbb{Z}(\bm{y})[x|x\in X]\subset \mathbb{Q}(\bm{y})(\bm{x}),
\end{equation}
where $X$ is the set of all the cluster variables obtained from iterative mutations to the initial seed.
\end{dfn}
Now we define the coefficient-free cluster algebra as the pair of a quiver and cluster variables $(Q,\bm{x})$, which is also called a seed, and by using its mutation.
\begin{dfn}
Let $\mu_k :(Q,\bm{x})\longmapsto(Q',\bm{x'})\quad (k=1,2,\dots ,N)$ be the mutation defined as follows.
\begin{itemize}
\item The definition of a new quiver $Q'$ is the same as that in Definition \ref{dfn:ca}.
\item Let $\bm{x'}=(x_1',x_2',\dots ,x_N')$ be the new cluster variables defined by $Q$ and $\bm{x}$ as:
\begin{equation}\label{clus}
\begin{aligned}
x_k'&=\frac{1}{x_k}\left( \prod_{\lambda_{k,j}>0}x_j^{\lambda_{k,j}}
+\prod_{\lambda_{j,k}>0}x_j^{\lambda_{j,k}}\right),\\
x_i'&=x_i\quad (i\neq k).
\end{aligned}
\end{equation}
\end{itemize}
\end{dfn}
\begin{dfn}
Let us fix an initial seed $(Q,\bm{x})$.
We define a coefficient-free cluster algebra $\mathcal{A}(Q,\bm{x})$ as
\begin{equation}
\mathcal{A}(Q,\bm{x})=\mathbb{Z}[x|x\in X]\subset \mathbb{Q}(\bm{x}),
\end{equation}
where $X$ is the set of cluster variables obtained from a iteration of all mutations to the initial seed.
\end{dfn}

\section{Bilinear equations satisfied by cluster variables and their reductions}

In this section, we show that cluster variables satisfy bilinear equations related to discrete integrable systems, if the initial seed includes suitable quivers with infinite vertices.
These quivers are obtained from a transformation of quivers which corresponds to a reduction of a certain difference equations.

\subsection{The discrete KdV equation and the Hirota-Miwa equation}

\begin{figure}
\begin{center}
\includegraphics[width=15cm]{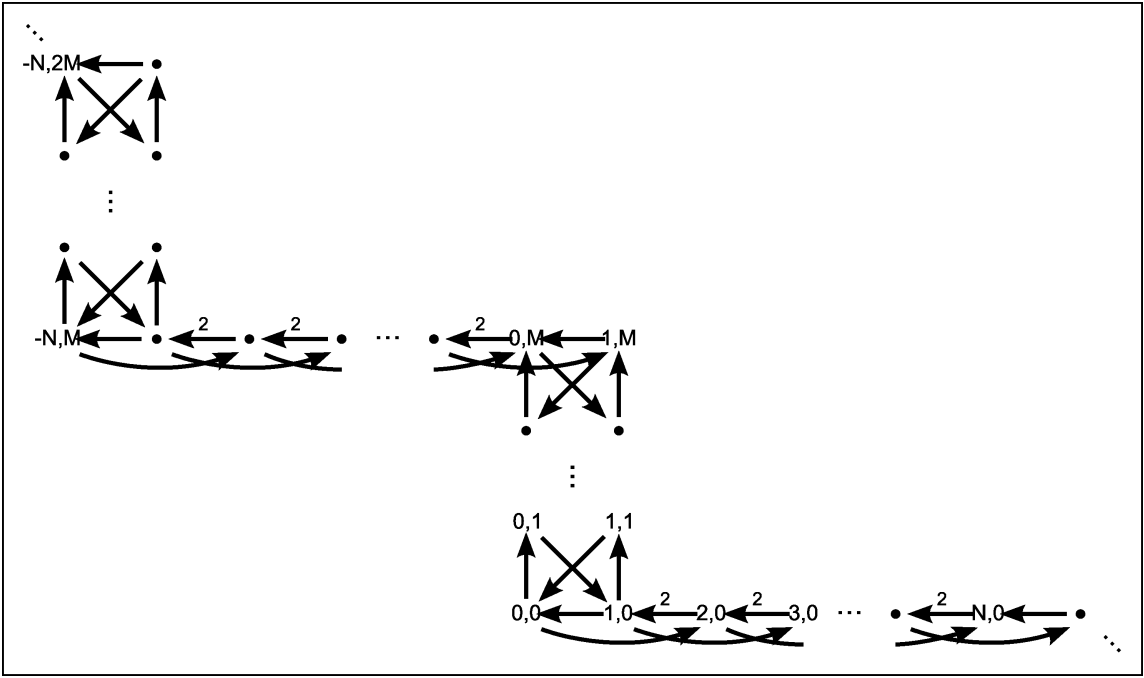}
\end{center}
\caption{The dKdV quiver (The numbers on vertices correspond to cluster variables.)}
\label{dKdV}
\end{figure}

\begin{figure}
\begin{center}
\includegraphics[width=15cm]{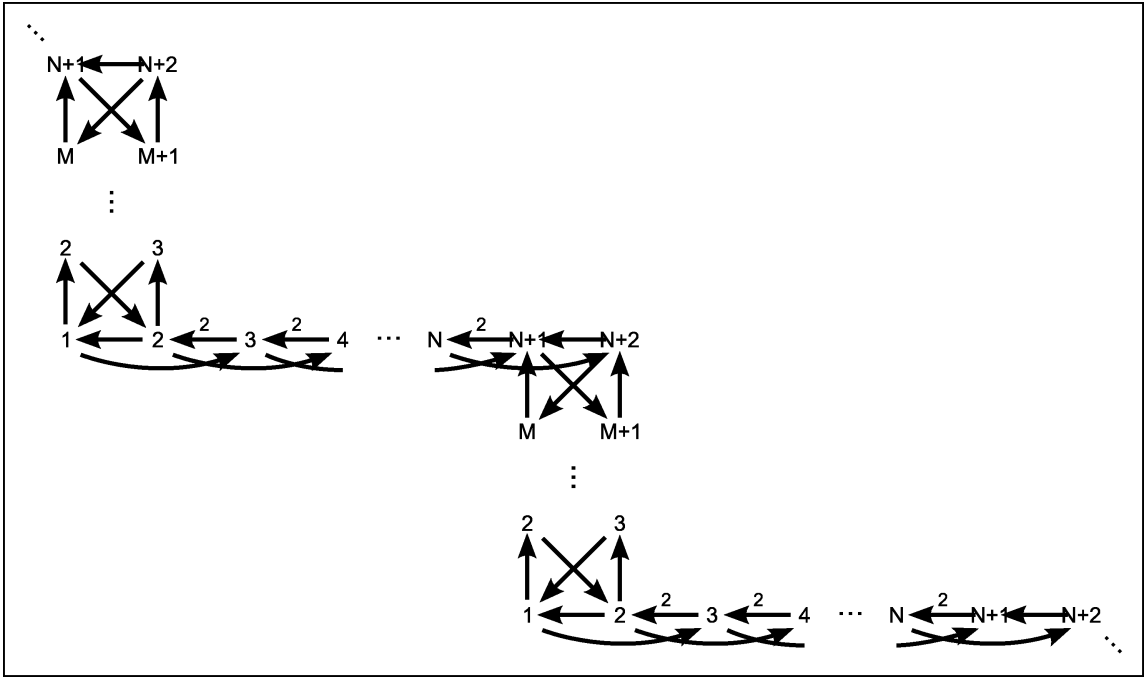}
\end{center}
\caption{The dKdV quiver (The numbers at the vertices denote the order of mutations.)}
\label{dKdVm}
\end{figure}

We construct cluster algebras whose cluster variables satisfy the discrete KdV equation and the Hirota-Miwa equation.
The quivers of the initial seeds were obtained in \cite{6}.
We consider quivers of general form of the discrete KdV equation and the Hirota-Miwa equation.
First we consider coefficient-free cluster algebras.
For any $N,M\geq1$, let $Q_{\rm{KdV}}^{N,M}$ be a quiver as shown in Figure \ref{dKdV} (the dKdV quiver), where the numbers attached to the arrows of the quivers denote the numbers of arrows pointing in the same direction.
Note that each vertex $(n,m)$ corresponds to a cluster variable $x_n^m$.
Let $\bm{x}$ be the set of these cluster variables.
We take $(Q_{\rm{KdV}}^{N,M},\bm{x})$ as an initial seed.
We define $\mu'_i$ as the iteration of all mutations at the vertices $i$ in Figure \ref{dKdVm}.
Figure \ref{dKdVm} shows the case where $N\geq M$.
Note that mutations at vertices with the same numbers are commutative.
We apply the mutation to the initial seed in the order $\mu'_1,\mu'_2,\dots,\mu'_{\max[N,M]},\mu'_1,\mu'_2,\dots$.
The new cluster variable obtained by mutation at $x_n^m$ is denoted by $x_{n+2}^{m+1}$.
We then obtain the following proposition from the definition of mutation \eqref{clus}.
\begin{prop}
Consider the coefficient-free cluster algebra $\mathcal{A}(Q_{\rm{KdV}}^{N,M},\bm{x})$.
For any $n,m\in \mathbb{Z}$, the cluster variables $x_n^m$ satisfy the bilinear equation
\begin{equation}\label{eq:dKdV}
x_{n+1}^{m+1}x_{n-1}^m=x_{n-1}^{m+1}x_{n+1}^m+x_n^{m+1}x_n^m.
\end{equation}
\end{prop}
This equation is nothing but the bilinear form of the discrete KdV equation \cite{4}.

\begin{figure}
\begin{center}
\includegraphics[width=15cm]{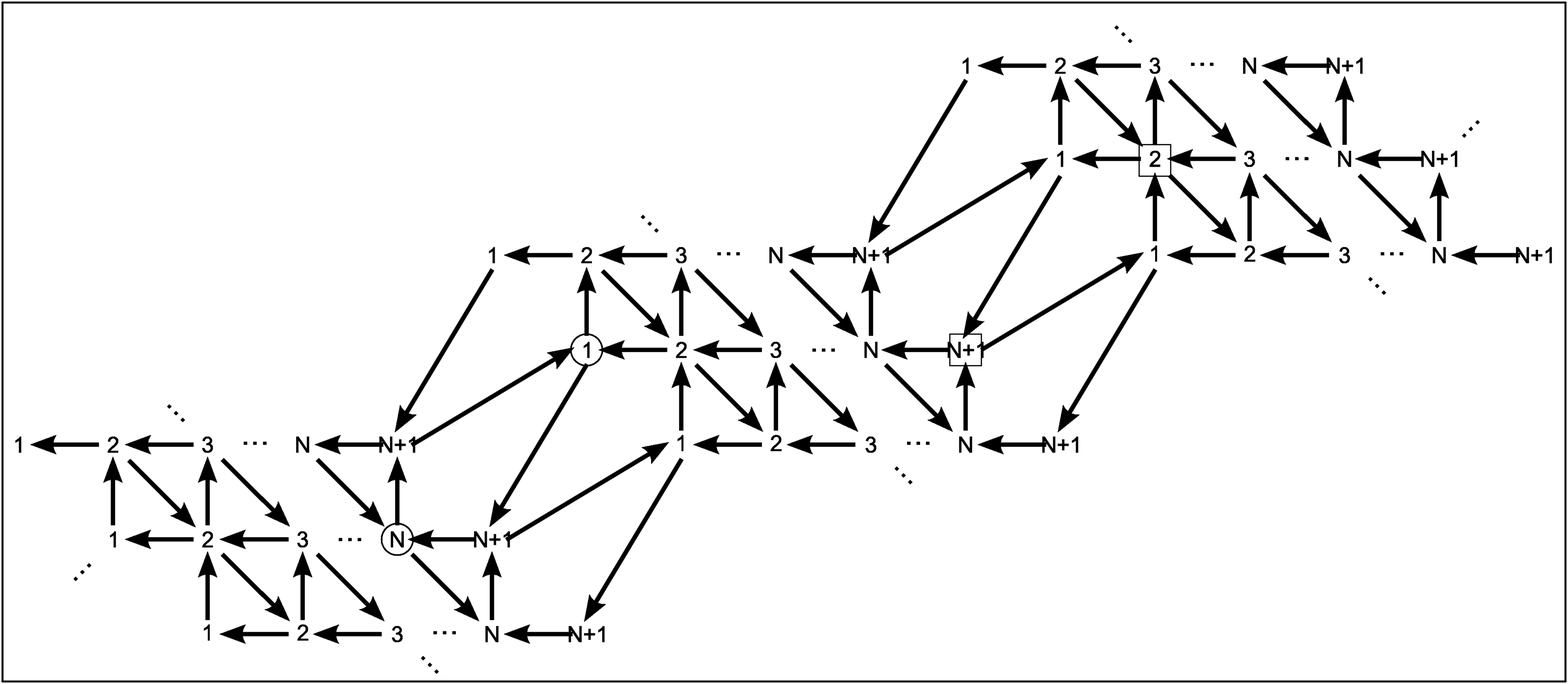}
\end{center}
\caption{The HM quiver (The numbers at the vertices denote the order of mutations. Vertices $1$ and $N$ contained in a circle correspond to the cluster variables $x_{0,0}^0$ and $x_{0,-1}^0$ respectively, and vertices $N+1$ and $2$ contained in a square correspond to the cluster variables $x_{N,0}^0$ and $x_{N,1}^0$ respectively.)}
\label{HM}
\end{figure}

For any $N\geq1$, let $Q_{\rm{HM}}^N$ be the quiver as shown in Figure \ref{HM} (the HM quiver).
Vertices $1$ and $N$ contained in a circle correspond to the cluster variables $x_{0,0}^0$ and $x_{0,-1}^0$ respectively, and vertices $N+1$ and $2$ contained in a square correspond to the cluster variables $x_{N,0}^0$ and $x_{N,1}^0$ respectively.
For the other vertices adjacent to $x_{n,l}^m$, we denote by $x_{n+1,l}^m$ the one on the right of  $x_{n,l}^m$, and by $x_{n,l}^{m+1}$ the one above $x_{n,l}^m$, and so on.
Let $\bm{x}$ be these cluster variables.
We take $(Q_{\rm{HM}}^N,\bm{x})$ as an initial seed and 
mutate it in the order $\mu'_1,\mu'_2,\dots,\mu'_{N+1},\mu'_1,\mu'_2,\dots$.
We denote by $x_{n+1,l-1}^{m+1}$ the new cluster variable obtained by mutation at $x_{n,l}^m$.
Then we obtain the following proposition from the definition of mutation \eqref{clus}.
\begin{prop}
Consider the coefficient-free cluster algebra $\mathcal{A}(Q_{\rm{HM}}^N,\bm{x})$.
For any $n,m,l\in \mathbb{Z}$, the cluster variables $x_{n,l}^m$ satisfy the bilinear equation
\begin{equation}\label{eq:HM}
x_{n+1,l}^{m+1}x_{n,l+1}^m
=x_{n,l+1}^{m+1}x_{n+1,l}^m+x_{n,l}^{m+1}x_{n+1,l+1}^m.
\end{equation}
\end{prop}
This equation is the Hirota-Miwa equation \cite{5}.

\begin{figure}
\begin{center}
\includegraphics[width=6cm]{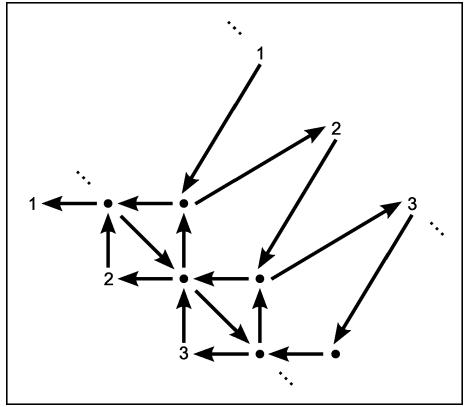}
\end{center}
\caption{Reduction from the HM quiver to the dKdV quiver (Superposition of vertices with same numbers.)}
\label{dKdVr}
\end{figure}

The quiver of the discrete KdV equation can be obtained from a certain transformation of the quiver of the Hirota-Miwa equation.
In particular, we show that the dKdV quiver $Q_{\rm{KdV}}^{1,1}$ is obtained from the HM quiver $Q_{\rm{HM}}^2$ by applying the following two operations successively on the quiver $Q_{\rm{HM}}^2$.
\begin{enumerate}
\item Among the arrows from $(n,m,l)+k(1,0,1)$ to $(n',m',l')+k(1,0,1)$ ($k\in\mathbb{Z}$),
we remove all the arrows with $k\neq 0$ in $Q_{\rm{HM}}^2$ (cf. Figure \ref{dKdVr}).
\item We superimpose the vertices $(n,m,l)+k(1,0,1)\ (k\in\mathbb{Z})$ on the vertex $(n,m,l)$. (In Figure \ref{dKdVr}, we superimpose the vertices with the same number.)
\end{enumerate}
The dKdV quiver $Q_{\rm{KdV}}^{1,1}$ is obtained from the above operation, which will be called a $(1,0,1)$-reduction of a quiver.
In a similar way the $(a,b,c)$-reduction of a quiver is defined.
In fact, reduction of a quiver corresponds to reduction of a difference equation.
In this case, the discrete KdV equation \eqref{eq:dKdV} is obtained from the Hirota-Miwa equation \eqref{eq:HM} by imposing the reduction condition $x_{n+1,l+1}^m=x_{n,l}^m$ and $x_n^m:=x_{n,0}^m$.

\subsection{The discrete mKdV equation and the discrete Toda equation}

\begin{figure}
\begin{minipage}{0.5\hsize}
\begin{center}
\includegraphics[width=7cm]{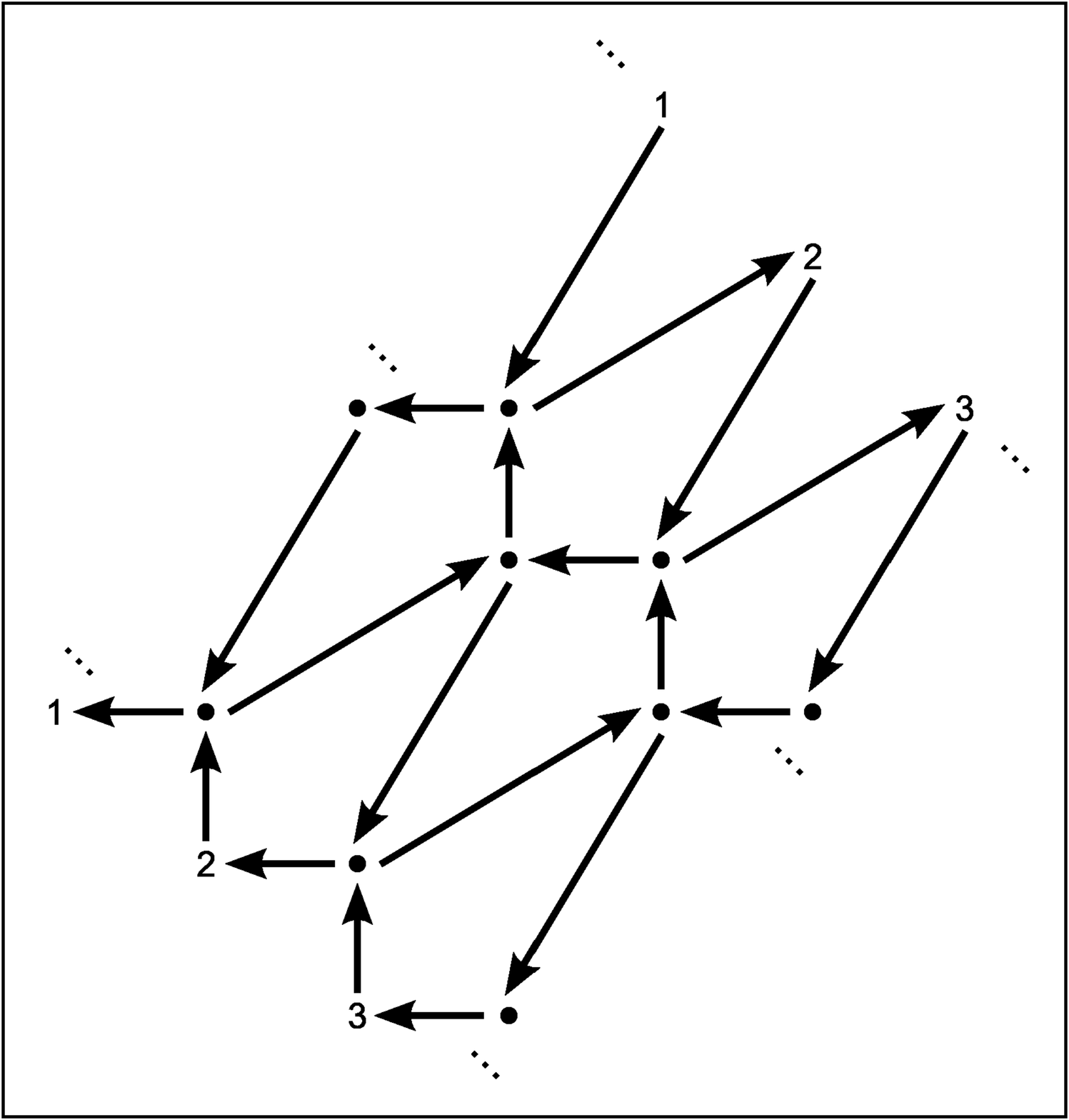}
\end{center}
\caption{Reduction from the HM quiver to the dmKdV quiver}
\label{dmKdVr}
\end{minipage}
\begin{minipage}{0.5\hsize}
\begin{center}
\includegraphics[width=7cm]{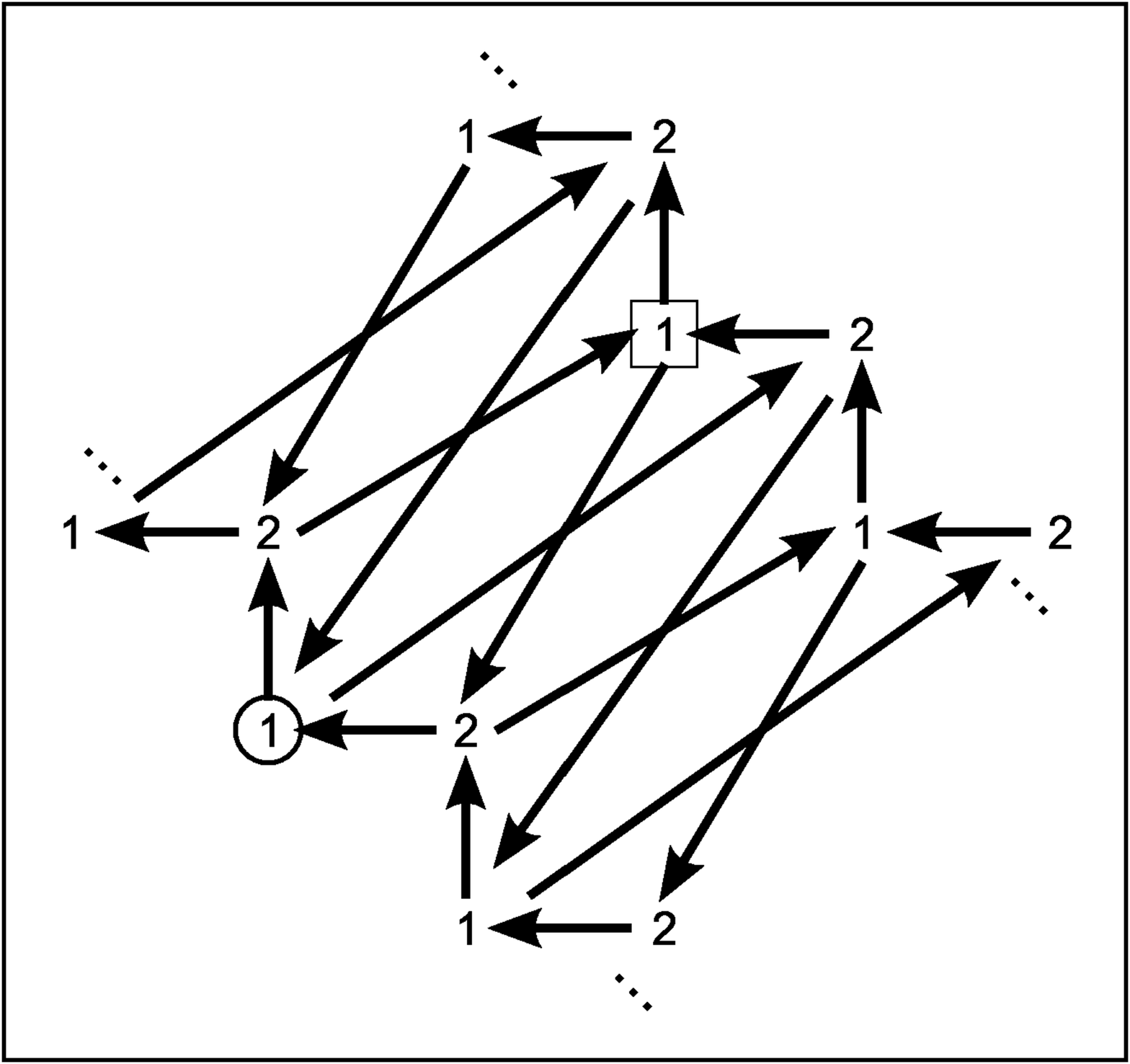}
\end{center}
\caption{The dmKdV quiver}
\label{dmKdV}
\end{minipage}
\end{figure}

We construct cluster algebras whose cluster variables satisfy the discrete mKdV equation and the discrete Toda equation by the reduction of the quiver of the Hirota-Miwa equation. 
Let $Q_{\rm{mKdV}}$ (the dmKdV quiver) be the quiver obtained from the $(0,0,2)$-reduction of the HM quiver $Q_{\rm{HM}}^1$ (cf. Figures \ref{dmKdVr}, \ref{dmKdV}).
In Figure \ref{dmKdV}, the numbers at vertices denote the order of mutations.
Vertices $1$ in a circle and a square correspond to the cluster variables $w_0^0$ and $x_0^0$ respectively.
For the other vertices adjacent to $w_n^m$ and $x_n^m$, we denote by $w_{n+1}^m$ and $x_{n+1}^m$ those on the right of $w_n^m,$ and $x_n^m$ respectively, and by $w_n^{m+1},$ and $x_n^{m+1}$ those above $w_n^m$ and $x_n^m$, and so on.
Let $\bm{x}$ be these cluster variables.
We take $(Q_{\rm{mKdV}},\bm{x})$ as an initial seed and mutate it in the order $\mu'_1,\mu'_2,\mu'_1,\mu'_2,\dots$.
We denote by $x_{n+1}^{m+1}$ and $w_{n+1}^{m+1}$ the new cluster variable obtained by mutation at $w_n^m$ and $x_n^m$ respectively.
Then we obtain the following proposition by the definition of mutation \eqref{clus}.
\begin{prop}
Consider the coefficient-free cluster algebra $\mathcal{A}(Q_{\rm{mKdV}},\bm{x})$.
For any $n,m\in \mathbb{Z}$, the cluster variables $w_n^m,x_n^m$ satisfy the bilinear equations:
\begin{equation}\label{eq:dmKdV}
\begin{aligned}
w_{n+1}^{m+1}x_n^m&=x_n^{m+1}w_{n+1}^m+w_n^{m+1}x_{n+1}^m,\\
x_{n+1}^{m+1}w_n^m&=w_n^{m+1}x_{n+1}^m+x_n^{m+1}w_{n+1}^m.
\end{aligned}
\end{equation}
\end{prop}
Note that nonautonomous bilinear equations can be obtained from cluster algebras with coefficients.
Consider the cluster algebra with coefficients $\mathcal{A}(Q_{\rm{mKdV}},\bm{x},\bm{y})$.
Let the cluster variables $w_n^m,x_n^m$ be defined as in the coefficient-free case.
By the definition of a mutation \eqref{clus2}, the cluster variables $w_n^m,x_n^m\ (n,m\in \mathbb{Z})$ satisfy the following bilinear equations:
\begin{equation}\label{eq:dmKdV2}
\begin{aligned}
w_{n+1}^{m+1}x_n^m&=a_n^mx_n^{m+1}w_{n+1}^m+b_n^mw_n^{m+1}x_{n+1}^m,\\
x_{n+1}^{m+1}w_n^m&=c_n^mw_n^{m+1}x_{n+1}^m+d_n^mx_n^{m+1}w_{n+1}^m,
\end{aligned}
\end{equation}
where $a_n^m,b_n^m,c_n^m,d_n^m$ are rational functions of the coefficients of the initial seed for which $a_n^m+b_n^m=c_n^m+d_n^m=1$ holds.
These equations are the bilinear form of the discrete mKdV equation.

\begin{figure}
\begin{center}
\includegraphics[width=9cm]{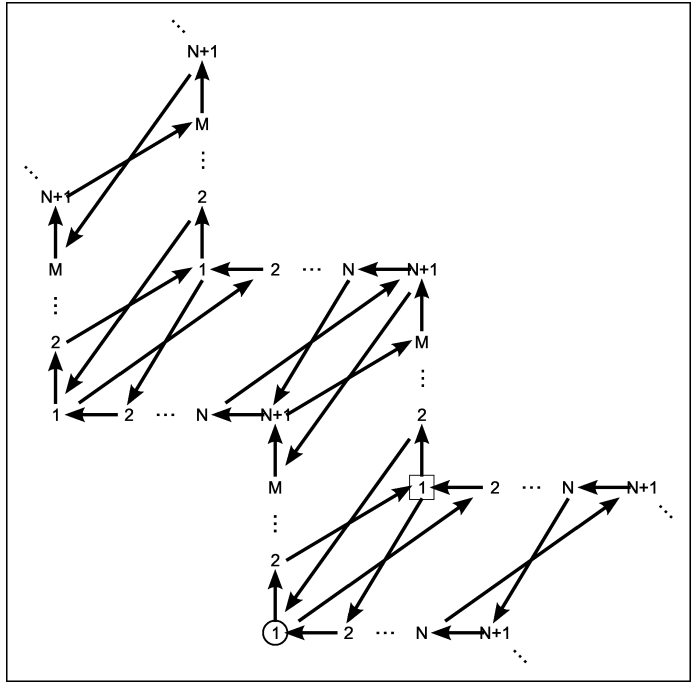}
\end{center}
\caption{Generalized dmKdV quiver}
\label{dmKdVg}
\end{figure}

The discrete mKdV equation \eqref{eq:dmKdV} is obtained from the Hirota-Miwa equation \eqref{eq:HM} by imposing the reduction condition $x_{n,l+2}^m=x_{n,l}^m$ and $w_n^m:=x_{n,0}^m,x_n^m:=x_{n,1}^m$.
This reduction of the difference equation corresponds to the $(0,0,2)$-reduction of the HM quiver.
Let $Q_{\rm{mKdV}}^{N,M}$ be the quiver as shown in Figure \ref{dmKdVg}, which shows the case of $N\geq M$.
This quiver is a generalization of the dmKdV quiver.
In fact, $Q_{\rm{mKdV}}=Q_{\rm{mKdV}}^{1,1}$.
In Figure \ref{dmKdVg}, numbers at vertices denote the order of mutations.
The discrete mKdV equations \eqref{eq:dmKdV} and \eqref{eq:dmKdV2} are obtained from the generalized dmKdV quiver in the same way.

\begin{figure}
\begin{minipage}{0.5\hsize}
\begin{center}
\includegraphics[width=7cm]{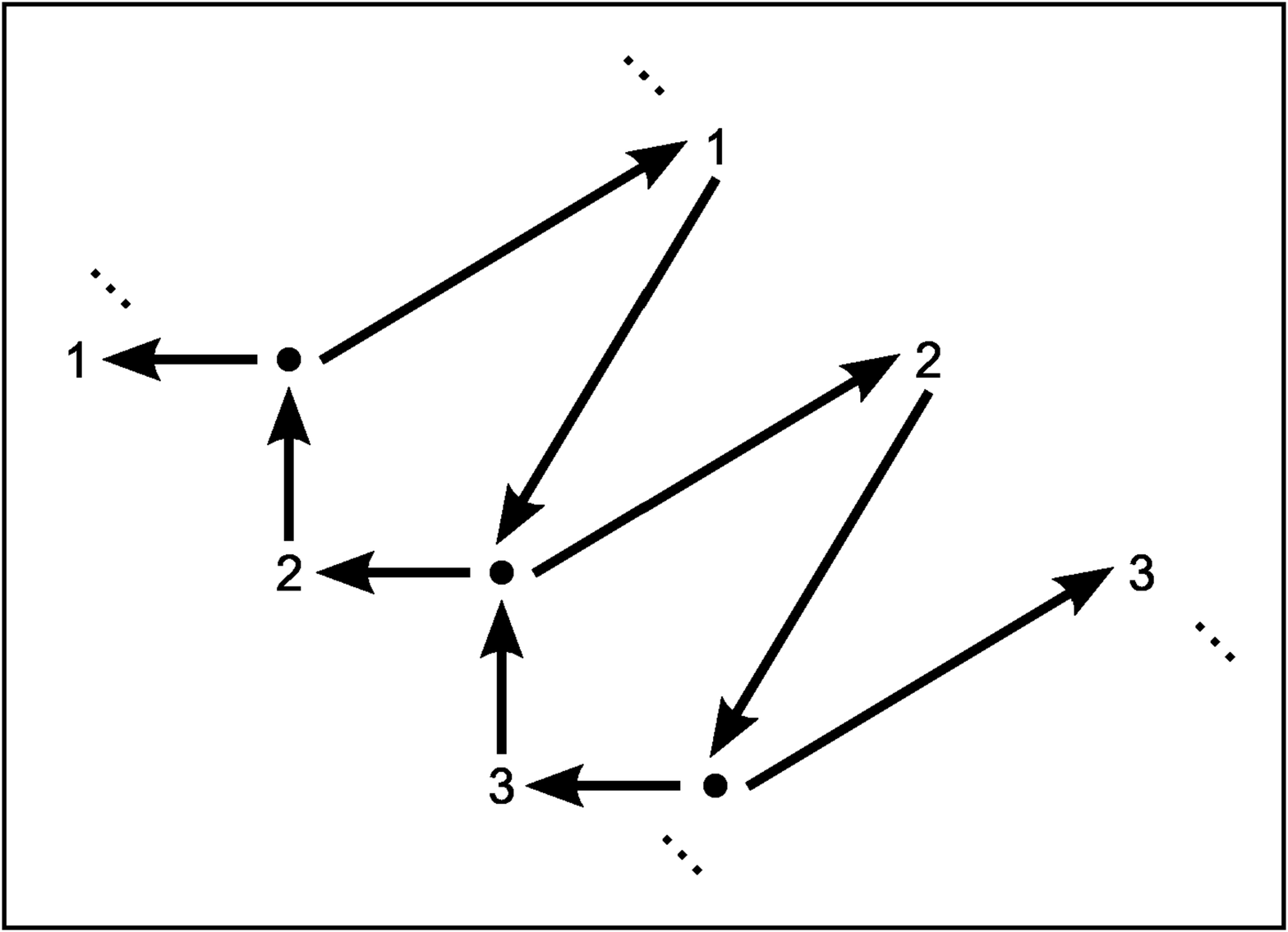}
\end{center}
\caption{Reduction from the HM quiver to the dToda quiver}
\label{dTodar}
\end{minipage}
\begin{minipage}{0.5\hsize}
\begin{center}
\includegraphics[width=5cm]{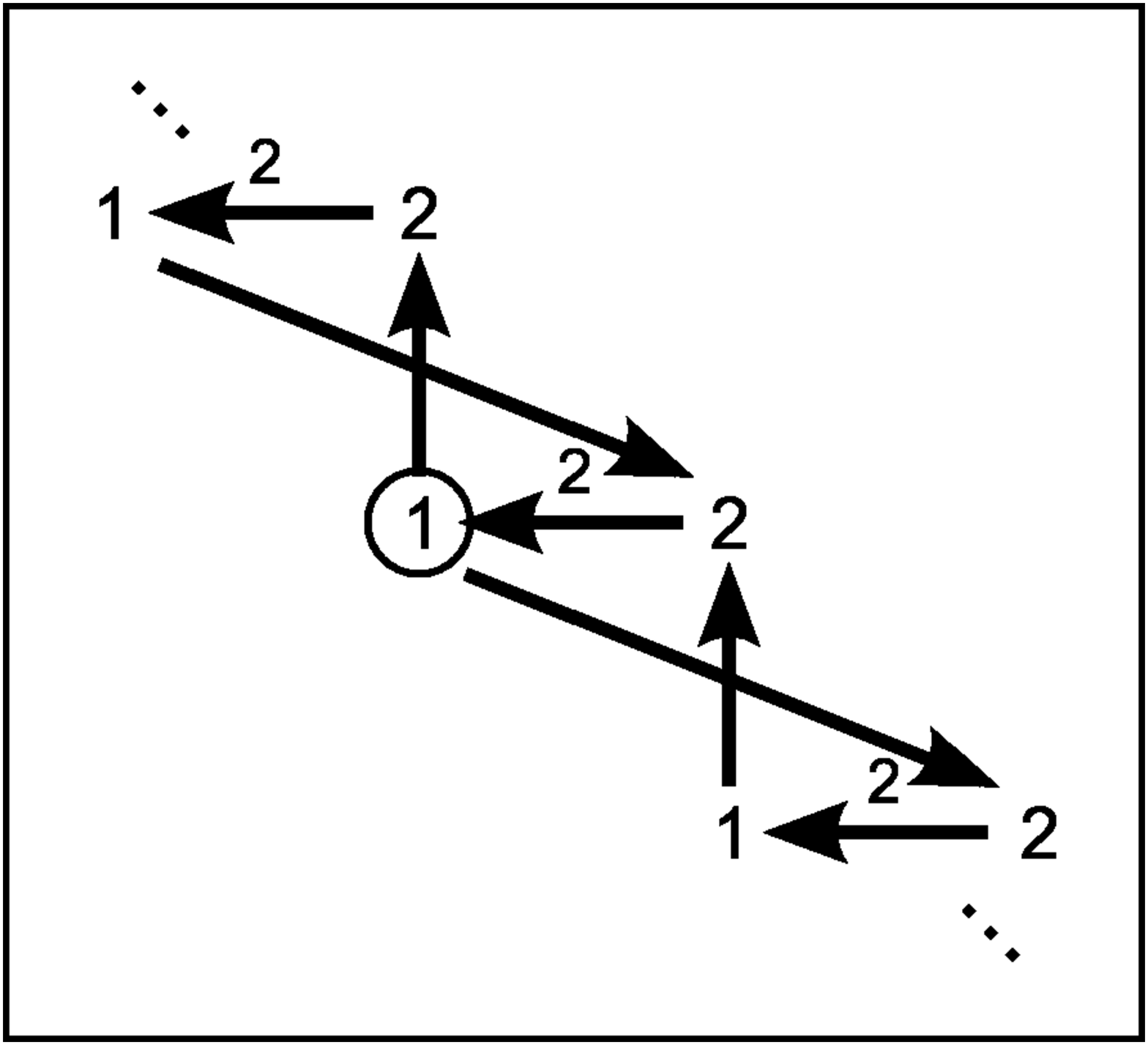}
\end{center}
\caption{The dToda quiver}
\label{dToda}
\end{minipage}
\end{figure}

Let $Q_{\rm{T}}$ (the dToda quiver) be the quiver obtained from the $(1,-1,1)$-reduction of the HM quiver $Q_{\rm{HM}}^1$ (cf. Figures \ref{dTodar}, \ref{dToda}).
In Figure \ref{dToda}, the numbers at vertices denote the order of mutations.
The vertex $1$ contained in a circle corresponds to the cluster variable $x_0^0$.
For the other vertices adjacent to $x_n^m$, we denote by $x_{n+1}^m$ the one on the right of $x_n^m$, and by $x_n^{m+1}$ that above $x_n^m$, and so on.
Let $\bm{x}$ be these cluster variables.
We take $(Q_{\rm{T}},\bm{x})$ as an initial seed and mutate it in the order $\mu'_1,\mu'_2,\mu'_1,\mu'_2,\dots$.
We denote by $x_{n+2}^m$ the new cluster variable obtained by mutation at $x_n^m$.
Then we obtain the following proposition by the definition of mutation \eqref{clus}.
\begin{prop}
Consider the coefficient-free cluster algebra $\mathcal{A}(Q_{\rm{T}},\bm{x})$.
For any $n,m\in \mathbb{Z}$, the cluster variables $x_n^m$ satisfy the bilinear equation
\begin{equation}\label{eq:dToda}
x_{n+1}^mx_{n-1}^m=x_{n-1}^{m+1}x_{n+1}^{m-1}+(x_n^m)^2.
\end{equation}
\end{prop}
This equation is the bilinear form of the discrete Toda equation.
The discrete Toda equation \eqref{eq:dToda} is obtained from the Hirota-Miwa equation \eqref{eq:HM} by imposing the reduction condition $x_{n+1,l+1}^{m-1}=x_{n,l}^m$ and $x_n^m:=x_{n,0}^m$.
This reduction of the difference equation corresponds to the $(1,-1,1)$-reduction of the HM quiver.

\section{$q$-discrete Painlev\'e equations satisfied by coefficients}

In this section, we show that coefficients in cluster algebras can satisfy $q$-discrete Painlev\'e equations, if the initial seed includes suitable quivers with the mutation-period property.
Quivers for the $q$-Painlev\'e I,II equations have been obtained in \cite{6,9}.
In this paper, we introduce the quivers for the $q$-Painlev\'e III,VI equations in a similar way.
We shall show that these quivers are obtained as reductions of the dKdV quiver and the dmKdV quiver.

\subsection{Mutation-periodic quivers}

The quivers of $q$-discrete Painlev\'e equations have the property that mutations of their quivers are equal to permutation of the vertices.
This is the so-called `mutation-period' property, which we define as follows.
Let $Q$ be a quiver.
For $\bm{i}=(i_1,i_2,\dots ,i_h)\ (i_j\in \{1,2,\dots ,N\})$, we define an iteration of mutations $\mu_{\bm{i}}$ as $\mu_{\bm{i}}(Q)=\mu_{i_h}\mu_{i_{h-1}}\cdots \mu_{i_1}(Q)$, where $h$ is the number of applications of the mutation.
For a permutation $\nu \in S_N$, let $\nu(Q)$ be the quiver in which we substituted vertices $i$ for $\nu(i)$ in $Q$.
\begin{dfn}
$\bm{i}$ is a $\nu$-period of $Q$ if $\mu_{\bm{i}}(Q)=\nu(Q)$ holds.
$Q$ is said to be a mutation-periodic quiver if $\bm{i}$ and $\nu \in S_N$, as defined above, exist.
\end{dfn}
In the case of $h=1\ (\bm{i}=(i_1))$, all mutation-periodic quivers have already been obtained \cite{3}.
Quivers of $q$-discrete Painlev\'e equations are mutation-periodic quivers.
These quivers arise from a reduction of the dKdV quiver and the dmKdV quiver.
We consider both cluster algebras with coefficients and coefficient-free cluster algebras.

\subsection{The $q$-Painlev\'e I equation}

\begin{figure}
\begin{minipage}{0.5\hsize}
\begin{center}
\includegraphics[width=5cm]{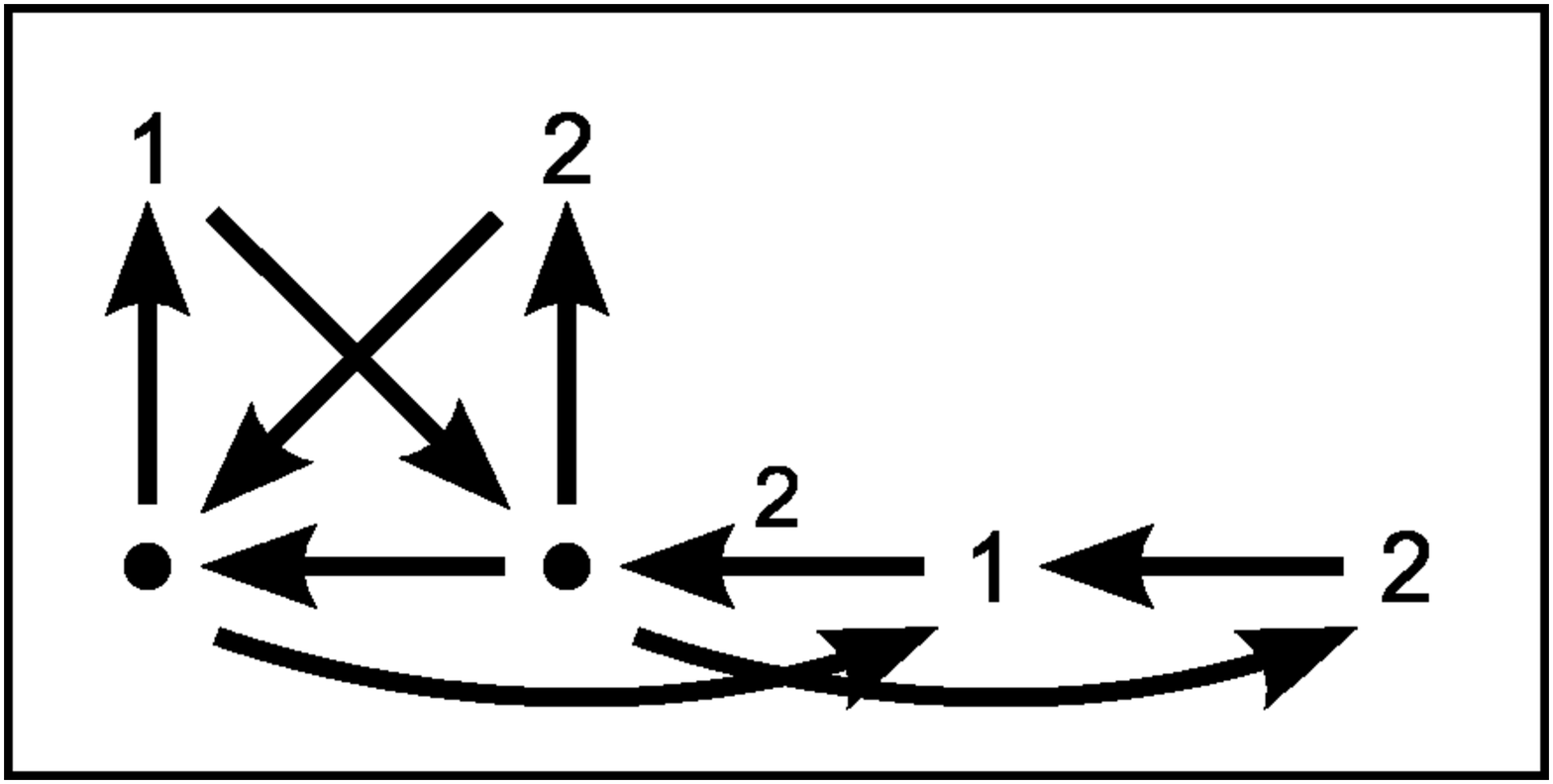}
\end{center}
\caption{Reduction from the dKdV quiver to the $q$-PI quiver}
\label{qPIr}
\end{minipage}
\begin{minipage}{0.5\hsize}
\begin{center}
\includegraphics[width=5cm]{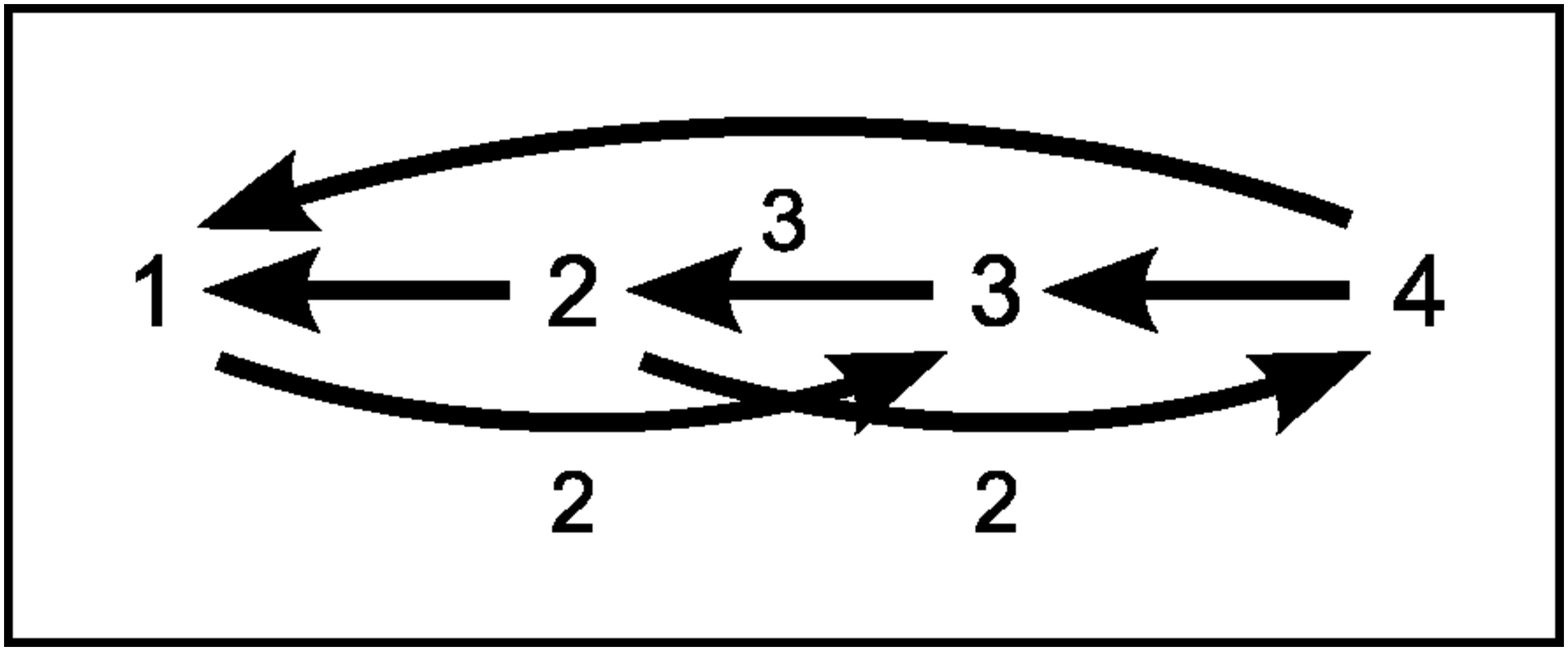}
\end{center}
\caption{The $q$-PI quiver}
\label{qPI}
\end{minipage}
\end{figure}

We construct the quiver of the $q$-Painlev\'e I equation by a reduction of the dKdV quiver.
Let $Q_{\rm{PI}}$ (the $q$-PI quiver) be the quiver obtained from the $(2,-1)$-reduction of the dKdV quiver $Q_{\rm{KdV}}^{2,1}$ (cf. Figures \ref{qPIr}, \ref{qPI}).
Let $\nu \in S_4$ be $\nu :(1,2,3,4)\mapsto (2,3,4,1)$.
$\bm{i}=(1)$ is a $\nu$-period of $Q_{\rm{PI}}$.
Note that each vertex $i$ corresponds to a cluster variable $x_i$ and a coefficient $y_{i,1}$.
We take $(Q_{\rm{PI}},\bm{x},\bm{y})$ as an initial seed, where $\bm{x}=(x_1,x_2,x_3,x_4),\bm{y}=(y_{1,1},y_{2,1},y_{3,1},y_{4,1})$ and we mutate the initial seed in the order $\mu_{\bm{i}}=\mu_1,\mu_{\nu(\bm{i})}=\mu_2,\mu_{\nu^2(\bm{i})}=\mu_3,\dots$.
The new cluster variables  and the new coefficients are denoted as $x_n\to x_{n+4},y_{n,m}\to y_{n,m+1}$.
We put $y_n:=y_{m,n}\ (n\equiv m\ (\rm{mod}\ 4))$ and obtain the following seeds:
\begin{equation}
\begin{aligned}
\cdots &\overset{\mu_4}{\longleftrightarrow}&
(Q_{\rm{PI}}&;x_1,x_2,x_3,x_4;y_1,y_{2,1},y_{3,1},y_{4,1})&\\
&\overset{\mu_1}{\longleftrightarrow}&
(\nu(Q_{\rm{PI}})&;x_5,x_2,x_3,x_4;y_{1,2},y_2,y_{3,2},y_{4,2})&\\
&\overset{\mu_2}{\longleftrightarrow}&
(\nu^2(Q_{\rm{PI}})&;x_5,x_6,x_3,x_4;y_{1,3},y_{2,3},y_3,y_{4,3})&\\
&\overset{\mu_3}{\longleftrightarrow}&
(\nu^3(Q_{\rm{PI}})&;x_5,x_6,x_7,x_4;y_{1,4},y_{2,4},y_{3,4},y_4)&\\
&\overset{\mu_4}{\longleftrightarrow}&
(Q_{\rm{PI}}&;x_5,x_6,x_7,x_8;y_5,y_{2,5},y_{3,5},y_{4,5})&\overset{\mu_1}{\longleftrightarrow}\cdots .
\end{aligned}
\end{equation}
The following proposition is obtained from the definition of mutation \eqref{clus}.
\begin{prop}
Consider the coefficient-free cluster algebra $\mathcal{A}(Q_{\rm{PI}},\bm{x})$.
For any $n\in \mathbb{Z}$, the cluster variables $x_n$ satisfy the bilinear equation
\begin{equation}\label{eq:qPIb}
x_{n+4}x_n=x_{n+2}^2+x_{n+3}x_{n+1}.
\end{equation}
\end{prop}
The bilinear equation \eqref{eq:qPIb} can be obtained from the discrete KdV equation \eqref{eq:dKdV} by imposing the reduction condition $x_{n+2}^{m-1}=x_n^m$ and $x_n:=x_n^0$.
This reduction of the difference equation corresponds to the $(2,-1)$-reduction of the dKdV quiver.
It turns out that the corresponding coefficients satisfy the $q$-Painlev\'e I equation.
\begin{thm}
Consider the cluster algebra with coefficient $\mathcal{A}(Q_{\rm{PI}},\bm{x},\bm{y})$.
For any $n\in \mathbb{Z}$, the coefficients $y_n$ satisfy the equation
\begin{equation}\label{eq:qPI}
y_{n+1}y_{n-1}=c_2c_1^n\frac{y_n+1}{y_n^2},
\end{equation}
where $c_1,c_2$ are the conserved quantities 
\begin{equation}
c_1=\frac{y_{n+3}\left(y_{n+1}^{-1}+1\right)}{y_n\left(y_{n+2}^{-1}+1\right)},\quad 
c_2=\frac{y_{n+2}y_{n+1}^2y_n}{y_{n+1}+1}c_1^{-(n+1)}
\end{equation}
and do not depend on $n$.
\end{thm}
This equation \eqref{eq:qPI} is the $q$-Painlev\'e I equation \cite{10}.
The $q$-Painlev\'e I equation \eqref{eq:qPI} is obtain from the bilinear equation \eqref{eq:qPIb} by the following transformation of variables:
\begin{equation}
y_n=\frac{x_{n+2}x_n}{x_{n+1}^2}.
\end{equation}
\begin{proof}
By the definition of mutation \eqref{coef}, the coefficients $y_{n,m}$ satisfy
\begin{equation}
\begin{aligned}
y_n&=y_{n,n-1}(y_{n-1}+1),\\
y_{n,n-1}&=y_{n,n-2}\left(y_{n-2}^{-1}+1\right)^{-2},\\
y_{n,n-2}&=y_{n,n-3}(y_{n-3}+1),\\
y_{n,n-3}&=y_{n-4}^{-1},
\end{aligned}
\end{equation}
where we consider the index $n$ of coefficients $y_{n,m}$ as $n\in\mathbb{Z}/4\mathbb{Z}$.
We then obtain an equation only for $y_n$:
\begin{equation}\label{eq:qPIy}
y_{n+4}=\frac{(y_{n+3}+1)(y_{n+1}+1)}{\left(y_{n+2}^{-1}+1\right)^2y_n}.
\end{equation}
We put 
\begin{equation}
u_n:=\frac{y_{n+3}\left(y_{n+1}^{-1}+1\right)}{y_n\left(y_{n+2}^{-1}+1\right)},\quad
v_n:=\frac{y_{n+2}y_{n+1}^2y_n}{y_{n+1}+1},
\end{equation}
and find $u_{n+1}=u_n$ from \eqref{eq:qPIy}.
Hence we obtain the conserved quantity $u_n=c_1$.
Similarly, we obtain $v_{n+1}=c_1v_n$ from $u_n=c_1$ and we obtain $v_n=c_2c_1^{n+1}$, where $c_2=v_nc_1^{-(n+1)}$ is also a conserved quantity.
We obtain the $q$-Painlev\'e I equation \eqref{eq:qPI} from $v_n=c_2c_1^{n+1}$.
\qed
\end{proof}

\subsection{The $q$-Painlev\'e II equation}

\begin{figure}
\begin{minipage}{0.5\hsize}
\begin{center}
\includegraphics[width=6cm]{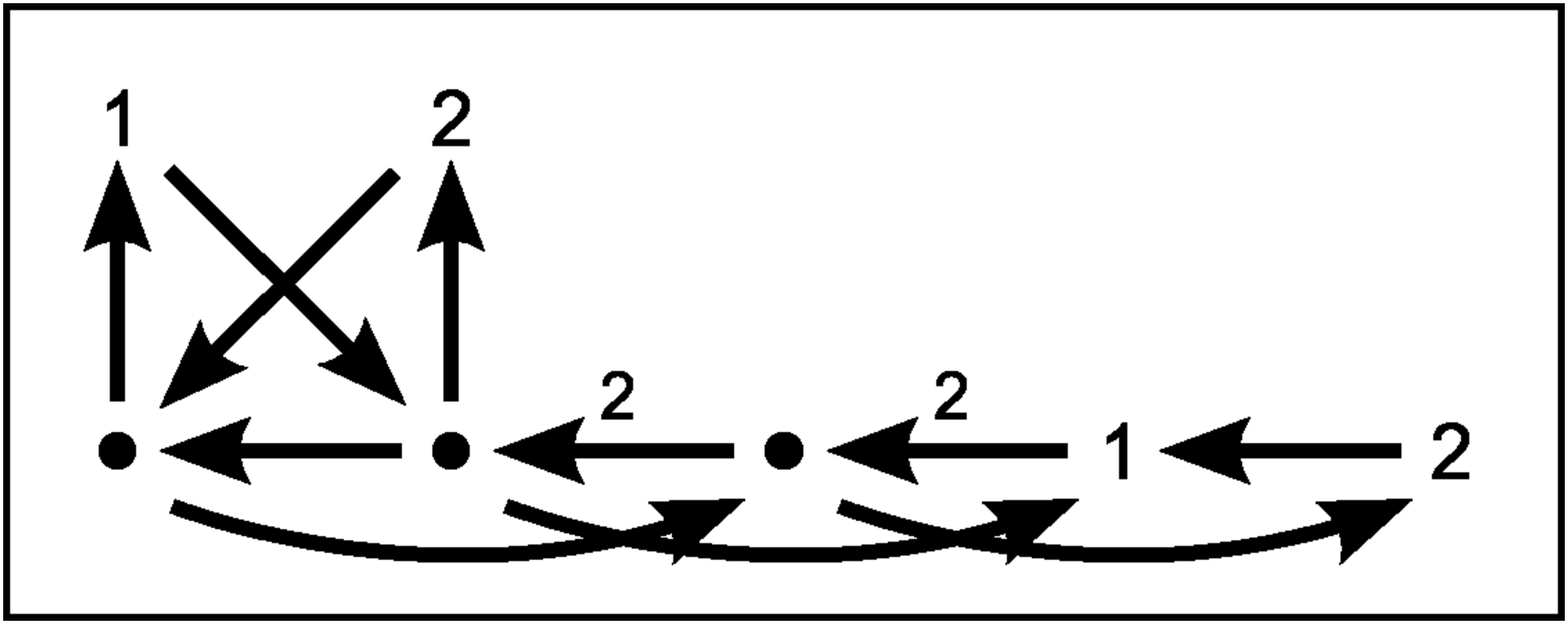}
\end{center}
\caption{Reduction from the dKdV quiver to the $q$-PII quiver}
\label{qPIIr}
\end{minipage}
\begin{minipage}{0.5\hsize}
\begin{center}
\includegraphics[width=6cm]{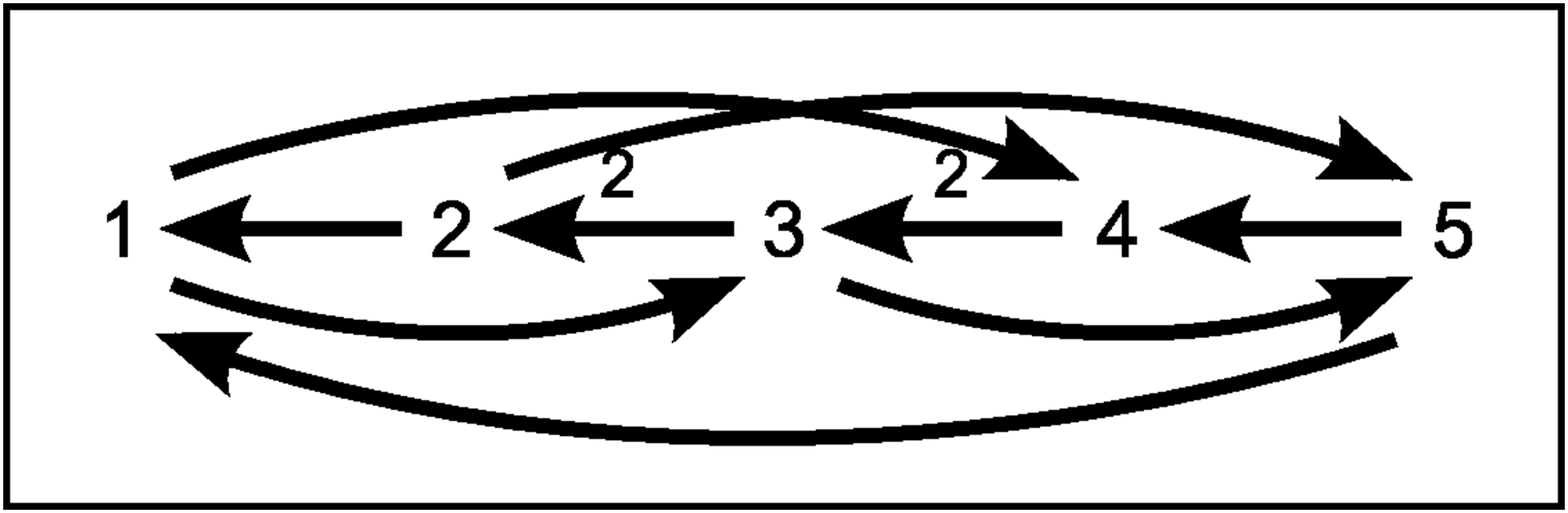}
\end{center}
\caption{The $q$-PII quiver}
\label{qPII}
\end{minipage}
\end{figure}

We construct the quiver of the $q$-Painlev\'e II equation by a reduction of dKdV quiver.
Let $Q_{\rm{PII}}$ (the $q$-PII quiver) be the quiver obtained from the $(3,-1)$-reduction of the dKdV quiver $Q_{\rm{KdV}}^{3,1}$ (cf. Figures \ref{qPIIr}, \ref{qPII}).
Let $\nu \in S_5$ be $\nu :(1,2,3,4,5)\mapsto (2,3,4,5,1)$.
$\bm{i}=(1)$ is a $\nu$-period of $Q_{\rm{PII}}$.
Note that each vertex $i$ corresponds to a cluster variable $x_i$ and a coefficient $y_{i,1}$.
We take $(Q_{\rm{PII}},\bm{x},\bm{y})$ as an initial seed, where $\bm{x}=(x_1,x_2,x_3,x_4,x_5),\bm{y}=(y_{1,1},y_{2,1},y_{3,1},y_{4,1},y_{5,1})$ and mutate the initial seed in the order $\mu_{\bm{i}},\mu_{\nu(\bm{i})},\mu_{\nu^2(\bm{i})},\dots$.
The new cluster variables and the new coefficients are denoted as $x_n\to x_{n+5},y_{n,m}\to y_{n,m+1}$.
We put $y_n:=y_{m,n}\ (n\equiv m\ (\rm{mod}\ 5))$ and obtain the following seeds:
\begin{equation}
\begin{aligned}
\cdots &\overset{\mu_5}{\longleftrightarrow}&
(Q_{\rm{PII}}&;x_1,x_2,x_3,x_4,x_5;
y_1,y_{2,1},y_{3,1},y_{4,1},y_{5,1})&\\
&\overset{\mu_1}{\longleftrightarrow}&
(\nu(Q_{\rm{PII}})&;x_6,x_2,x_3,x_4,x_5;
y_{1,2},y_2,y_{3,2},y_{4,2},y_{5,2})&\\
&\overset{\mu_2}{\longleftrightarrow}&
(\nu^2(Q_{\rm{PII}})&;x_6,x_7,x_3,x_4,x_5;
y_{1,3},y_{2,3},y_3,y_{4,3},y_{5,3})&
\overset{\mu_3}{\longleftrightarrow}\cdots.
\end{aligned}
\end{equation}
We obtain the following proposition from the definition of mutation \eqref{clus}.
\begin{prop}
Consider the coefficient-free cluster algebra $\mathcal{A}(Q_{\rm{PII}},\bm{x})$.
For any $n\in \mathbb{Z}$, the cluster variables $x_n$ satisfy the bilinear equation
\begin{equation}\label{eq:qPIIb}
x_{n+5}x_n=x_{n+3}x_{n+2}+x_{n+4}x_{n+1}.
\end{equation}
\end{prop}
The bilinear equation \eqref{eq:qPIIb} is obtained from the discrete KdV equation \eqref{eq:dKdV} by imposing the reduction condition $x_{n+3}^{m-1}=x_n^m$ and $x_n:=x_n^0$.
This reduction of the difference equation corresponds to the $(3,-1)$-reduction of the dKdV quiver.
Its coefficients satisfy the $q$-Painlev\'e II equation.
\begin{thm}
Consider the cluster algebra with coefficient $\mathcal{A}(Q_{\rm{PII}},\bm{x},\bm{y})$.
For any $n\in \mathbb{Z}$, the coefficients $y_n$ satisfy the equation
\begin{equation}\label{eq:qPII}
y_{n+1}y_{n-1}=c_2c_3^{(-1)^n}c_1^n\frac{y_n+1}{y_n},
\end{equation}
where $c_1,c_2,c_3$ are the conserved quantities 
\begin{equation}
\begin{aligned}
c_1^2&=\frac{y_{n+4}\left(y_{n+1}^{-1}+1\right)}{y_n\left(y_{n+3}^{-1}+1\right)},\quad 
c_2^2=\frac{y_{2n+3}y_{2n+2}^2y_{2n+1}^2y_{2n}}{(y_{2n+2}+1)(y_{2n+1}+1)}c_1^{-(4n+3)},\\
c_3^2&=\frac{y_{2n+3}(y_{2n+1}+1)}{y_{2n}(y_{2n+2}+1)}c_1^{-1}
\end{aligned}
\end{equation}
and do not depend on $n$.
\end{thm}
We put $f_n:=y_{2n},g_n:=y_{2n+1}$ and obtain 
\begin{equation}
\begin{aligned}
f_{n+1}f_n&=c_2c_3^{-1}c_1^{2n+1}\frac{g_n+1}{g_n},\\
g_ng_{n-1}&=c_2c_3c_1^{2n}\frac{f_n+1}{f_n}
\end{aligned}
\end{equation}
from \eqref{eq:qPII}.
This equation is the $q$-Painlev\'e II equation \cite{11}.
The $q$-Painlev\'e II equation \eqref{eq:qPII} is obtained from the bilinear equation \eqref{eq:qPIIb} by the following transformation of variables:
\begin{equation}
y_n=\frac{x_{n+3}x_n}{x_{n+2}x_{n+1}}.
\end{equation}
\begin{proof}
By the definition of mutation \eqref{coef}, the coefficients $y_{n,m}$ satisfy
\begin{equation}
\begin{aligned}
y_n&=y_{n,n-1}(y_{n-1}+1),\\
y_{n,n-1}&=y_{n,n-2}\left(y_{n-2}^{-1}+1\right)^{-1},\\
y_{n,n-2}&=y_{n,n-3}\left(y_{n-3}^{-1}+1\right)^{-1},\\
y_{n,n-3}&=y_{n,n-4}(y_{n-4}+1),\\
y_{n,n-4}&=y_{n-5}^{-1},
\end{aligned}
\end{equation}
where we think of the index $n$ of the coefficients $y_{n,m}$ as $n\in\mathbb{Z}/5\mathbb{Z}$.
We obtain an equation only for $y_n$:
\begin{equation}\label{eq:qPIIy}
y_{n+5}=\frac{(y_{n+4}+1)(y_{n+1}+1)}{\left(y_{n+3}^{-1}+1\right)\left(y_{n+2}^{-1}+1\right)y_n}.
\end{equation}
We put
\begin{equation}
u_n:=\frac{y_{n+4}\left(y_{n+1}^{-1}+1\right)}{y_n\left(y_{n+3}^{-1}+1\right)},\quad
v_n:=\frac{y_{n+2}y_n}{y_{n+1}^{-1}+1},
\end{equation}
and find $u_{n+1}=u_n$ from \eqref{eq:qPIIy}.
Hence, we obtain the conserved quantity $u_n=c_1^2$.
Similarly, we obtain $v_{n+2}=c_1^2v_n$ from $u_n=c_1^2$, and $v_n=c_2c_3^{(-1)^{n+1}}c_1^{n+1}$; $c_2^2=v_{2n+1}v_{2n}c_1^{-(4n+3)}$ and $c_3^2=v_{2n+1}v_{2n}^{-1}c_1^{-1}$ are conserved quantities.
We then obtain the $q$-Painlev\'e II equation \eqref{eq:qPII} from $v_n=c_2c_3^{(-1)^{n+1}}c_1^{n+1}$.
\qed
\end{proof}

\subsection{The $q$-Painlev\'e III equation}

\begin{figure}
\begin{minipage}{0.5\hsize}
\begin{center}
\includegraphics[width=5cm]{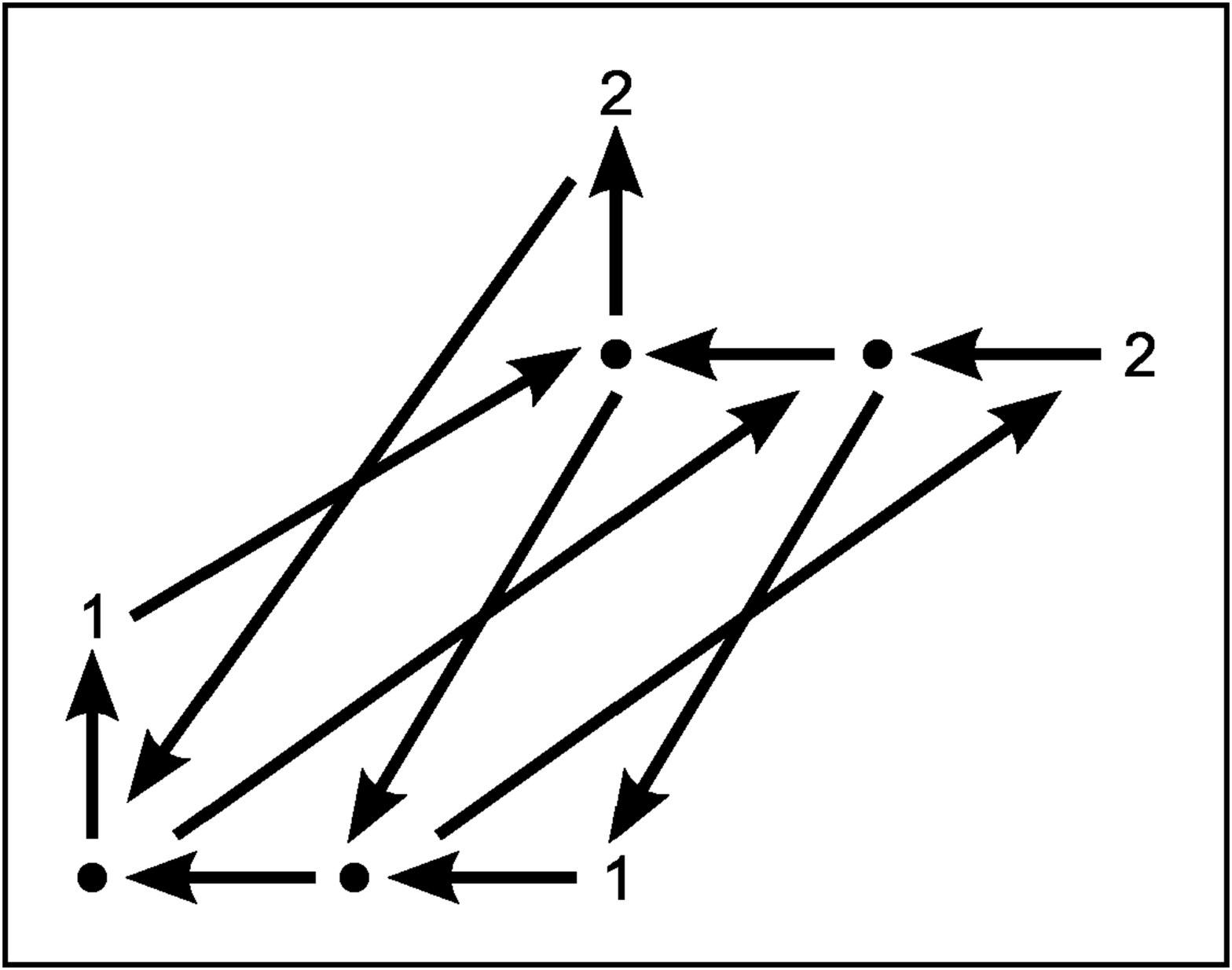}
\end{center}
\caption{Reduction from the dmKdV quiver to the $q$-PIII quiver}
\label{qPIIIr}
\end{minipage}
\begin{minipage}{0.5\hsize}
\begin{center}
\includegraphics[width=5cm]{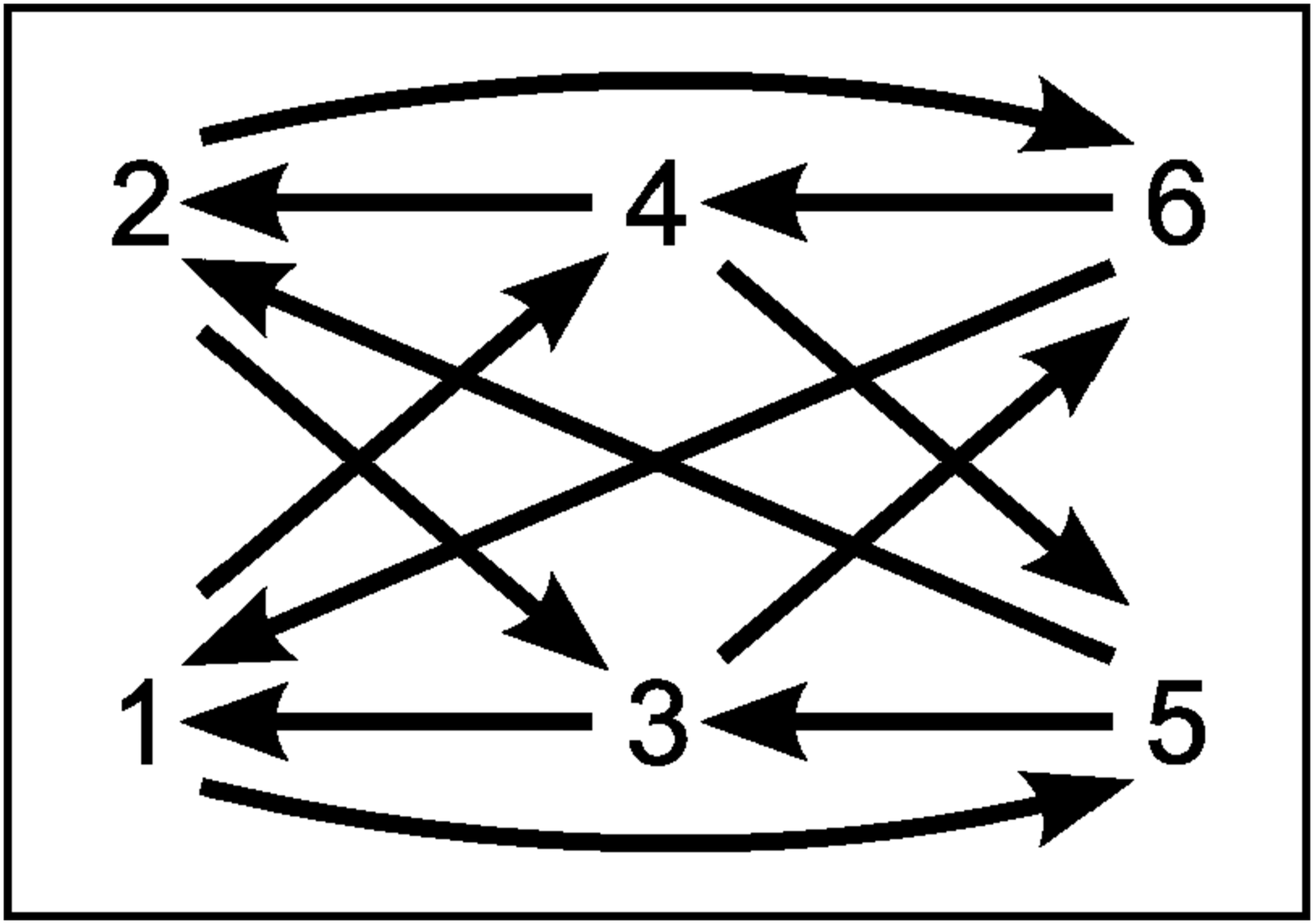}
\end{center}
\caption{The $q$-PIII quiver}
\label{qPIII}
\end{minipage}
\end{figure}

We construct the quiver of the $q$-Painlev\'e III equation by a reduction of the dmKdV quiver.
Let $Q_{\rm{PIII}}$ (the $q$-PIII quiver) be the quiver obtained from the $(2,-1)$-reduction of the dmKdV quiver $Q_{\rm{mKdV}}^{2,1}$ (cf. Figures \ref{qPIIIr}, \ref{qPIII}).
Let $\nu \in S_6$ be $\nu :(1,2,3,4,5,6)\mapsto (3,4,5,6,2,1)$.
$\bm{i}=(1,2)$ is a $\nu$-period of $Q_{\rm{PIII}}$.
Note that vertices $(1,2,3,4,5,6)$ correspond to cluster variables $\bm{x}=(w_1,x_1,w_2,x_2,w_3,x_3)$ and coefficients $\bm{y}=(y_{1,1},z_{1,1},y_{2,1},z_{2,1},y_{3,1},z_{3,1})$.
We take $(Q_{\rm{PIII}},\bm{x},\bm{y})$ as an initial seed and mutate it in the order $\mu_{\bm{i}}=\mu_2\mu_1,\mu_{\nu(\bm{i})}=\mu_4\mu_3,\mu_{\nu^2(\bm{i})}=\mu_6\mu_5,\dots$.
The new cluster variables are denoted as $w_n\to x_{n+3},\ x_n\to w_{n+3}$ and the new coefficients are denoted as $\ y_{n,m}\to z_{n,m+1},\ z_{n,m}\to y_{n,m+1}$ (if the coefficient corresponds to the vertex to which the mutation is applied) or $\ y_{n,m}\to y_{n,m+1},\ z_{n,m}\to z_{n,m+1}$ (otherwise).
We put $y_n:=y_{m,n},z_n:=z_{m,n}\ (n\equiv m\ (\rm{mod}\ 3))$ and obtain the following seeds:
\begin{equation}
\begin{aligned}
\cdots &\overset{\mu_5\mu_6}{\longleftrightarrow}&
(Q_{\rm{PIII}}&;w_1,x_1,w_2,x_2,w_3,x_3;
y_1,z_1,y_{2,1},z_{2,1},y_{3,1},z_{3,1})&\\
&\overset{\mu_2\mu_1}{\longleftrightarrow}&
(\nu(Q_{\rm{PIII}})&;x_4,w_4,w_2,x_2,w_3,x_3;
z_{1,2},y_{1,2},y_2,z_2,y_{3,2},z_{3,2})&\\
&\overset{\mu_4\mu_3}{\longleftrightarrow}&
(\nu^2(Q_{\rm{PIII}})&;x_4,w_4,x_5,w_5,w_3,x_3;
z_{1,3},y_{1,3},z_{2,3},y_{2,3},y_3,z_3)&\\
&\overset{\mu_6\mu_5}{\longleftrightarrow}&
(Q_{\rm{PIII}}&;x_4,w_4,x_5,w_5,x_6,w_6;
z_4,y_4,z_{2,4},y_{2,4},z_{3,4},y_{3,4})&\\
&\overset{\mu_1\mu_2}{\longleftrightarrow}&
(\nu(Q_{\rm{III}})&;w_7,x_7,x_5,w_5,x_6,w_6;
y_{1,5},z_{1,5},z_5,y_5,z_{3,5},y_{3,5})&\overset{\mu_3\mu_4}{\longleftrightarrow}\cdots.
\end{aligned}
\end{equation}
We obtain the following proposition from the definition of mutation \eqref{clus}.
\begin{prop}
Consider the coefficient-free cluster algebra $\mathcal{A}(Q_{\rm{PIII}},\bm{x})$.
For any $n\in \mathbb{Z}$, the cluster variables $w_n,x_n$ satisfy the bilinear equations:
\begin{equation}\label{eq:qPIIIb}
\begin{aligned}
w_{n+3}x_n&=x_{n+2}w_{n+1}+w_{n+2}x_{n+1},\\
x_{n+3}w_n&=w_{n+2}x_{n+1}+x_{n+2}w_{n+1}.
\end{aligned}
\end{equation}
\end{prop}
The bilinear equations \eqref{eq:qPIIIb} are obtained from the discrete mKdV equation \eqref{eq:dmKdV} by imposing the reduction condition $w_{n+2}^{m-1}=w_n^m,x_{n+2}^{m-1}=x_n^m$ and $w_n:=w_n^0,x_n:=x_n^0$.
This reduction of the difference equation corresponds to the $(2,-1)$-reduction of the dmKdV quiver.
In this case, the coefficients satisfy the $q$-Painlev\'e III equation.
\begin{thm}
Consider the cluster algebra with coefficient $\mathcal{A}(Q_{\rm{PIII}},\bm{x},\bm{y})$.
For any $n\in \mathbb{Z}$, the coefficients $y_n,z_n$ satisfy the equations:
\begin{equation}\label{eq:qPIII}
\begin{aligned}
y_{n+1}y_{n-1}&=c_2c_3^2c_1^{2n}\frac{y_n+1}{y_n\left(y_n+c_3c_4^{(-1)^n}c_1^n\right)},\\
z_{n+1}z_{n-1}&=c_2^{-1}c_3^2c_1^{2n}\frac{z_n+1}{z_n\left(z_n+c_3c_4^{(-1)^n}c_1^n\right)},
\end{aligned}
\end{equation}
where $c_1,c_2,c_3,c_4$ are conserved quantities
\begin{equation}
\begin{aligned}
&c_1^2=\frac{y_{n+2}z_{n+2}}{y_nz_n},\quad
&c_2^2&=\frac{y_{n+2}y_n\left(z_{n+1}^{-1}+1\right)^2}{z_{n+2}z_n\left(y_{n+1}^{-1}+1\right)^2},\\
&c_3^2=y_{2n+1}z_{2n+1}y_{2n}z_{2n}c_1^{-(4n+1)},\quad
&c_4^2&=\frac{y_{2n}z_{2n}}{y_{2n+1}z_{2n+1}}c_1
\end{aligned}
\end{equation}
and do not depend on $n$.
\end{thm}
If we put $f_n:=y_{2n},g_n:=y_{2n+1}$, we obtain 
\begin{equation}
\begin{aligned}
f_{n+1}f_n&=c_2c_3^2c_1^{4n+2}\frac{g_n+1}{g_n\left(g_n+c_3c_4^{-1}c_1^{2n+1}\right)},\\
g_ng_{n-1}&=c_2c_3^2c_1^{4n}\frac{f_n+1}{f_n\left(f_n+c_3c_4c_1^{2n}\right)}
\end{aligned}
\end{equation}
from \eqref{eq:qPIII}.
These equations are the $q$-Painlev\'e III equation \cite{12}.
\begin{proof}
By the definition of mutation \eqref{coef}, the coefficients $y_{n,m},z_{n,m}$ satisfy
\begin{equation}
\begin{aligned}
y_n&=y_{n,n-1}(y_{n-1}+1)\left(z_{n-1}^{-1}+1\right)^{-1},\\
y_{n,n-1}&=y_{n,n-2}\left(y_{n-2}^{-1}+1\right)^{-1}(z_{n-2}+1),\\
y_{n,n-2}&=z_{n-3}^{-1},\\
z_n&=z_{n,n-1}(z_{n-1}+1)\left(y_{n-1}^{-1}+1\right)^{-1},\\
z_{n,n-1}&=z_{n,n-2}\left(z_{n-2}^{-1}+1\right)^{-1}(y_{n-2}+1),\\
z_{n,n-2}&=y_{n-3}^{-1},
\end{aligned}
\end{equation}
where we consider the index $n$ of the coefficients $y_{n,m},z_{n,m}$ as $n\in\mathbb{Z}/3\mathbb{Z}$.
The equations, only for $y_n,z_n$, are
\begin{equation}\label{eq:qPIIIy}
\begin{aligned}
y_{n+3}&=\frac{(y_{n+2}+1)(z_{n+1}+1)}{\left(z_{n+2}^{-1}+1\right)\left(y_{n+1}^{-1}+1\right)z_n},\\
z_{n+3}&=\frac{(z_{n+2}+1)(y_{n+1}+1)}{\left(y_{n+2}^{-1}+1\right)\left(z_{n+1}^{-1}+1\right)y_n}.
\end{aligned}
\end{equation}
If we put
\begin{equation}
u_n:=\frac{y_{n+2}\left(z_{n+1}^{-1}+1\right)}{z_n\left(y_{n+1}^{-1}+1\right)},\quad
v_n:=\frac{z_{n+2}\left(y_{n+1}^{-1}+1\right)}{y_n\left(z_{n+1}^{-1}+1\right)},\quad 
t_n:=y_nz_n,
\end{equation}
we find $u_{n+1}=u_n,v_{n+1}=v_n$ from \eqref{eq:qPIIIy}, and obtain conserved quantities $u_n=c_1c_2,v_n=c_1c_2^{-1}$.
We also obtain $t_{n+2}=c_1^2t_n$ from $u_nv_n=c_1^2$, and $t_n=c_3c_4^{(-1)^n}c_1^n$.
$c_3^2=t_{2n+1}t_{2n}c_1^{-(4n+1)}$ and $c_4^2=t_{2n+1}^{-1}t_{2n}c_1$ are conserved quantities.
We obtain the $q$-Painlev\'e III equation \eqref{eq:qPIII} from $u_n=c_1c_2$ and $t_n=c_3c_4^{(-1)^n}c_1^n$ by the elimination of $z_n$.
The equation for $z_n$ is the same as that for $y_n$.
\qed
\end{proof}

\subsection{The $q$-Painlev\'e VI equation}

\begin{figure}
\begin{minipage}{0.5\hsize}
\begin{center}
\includegraphics[width=5cm]{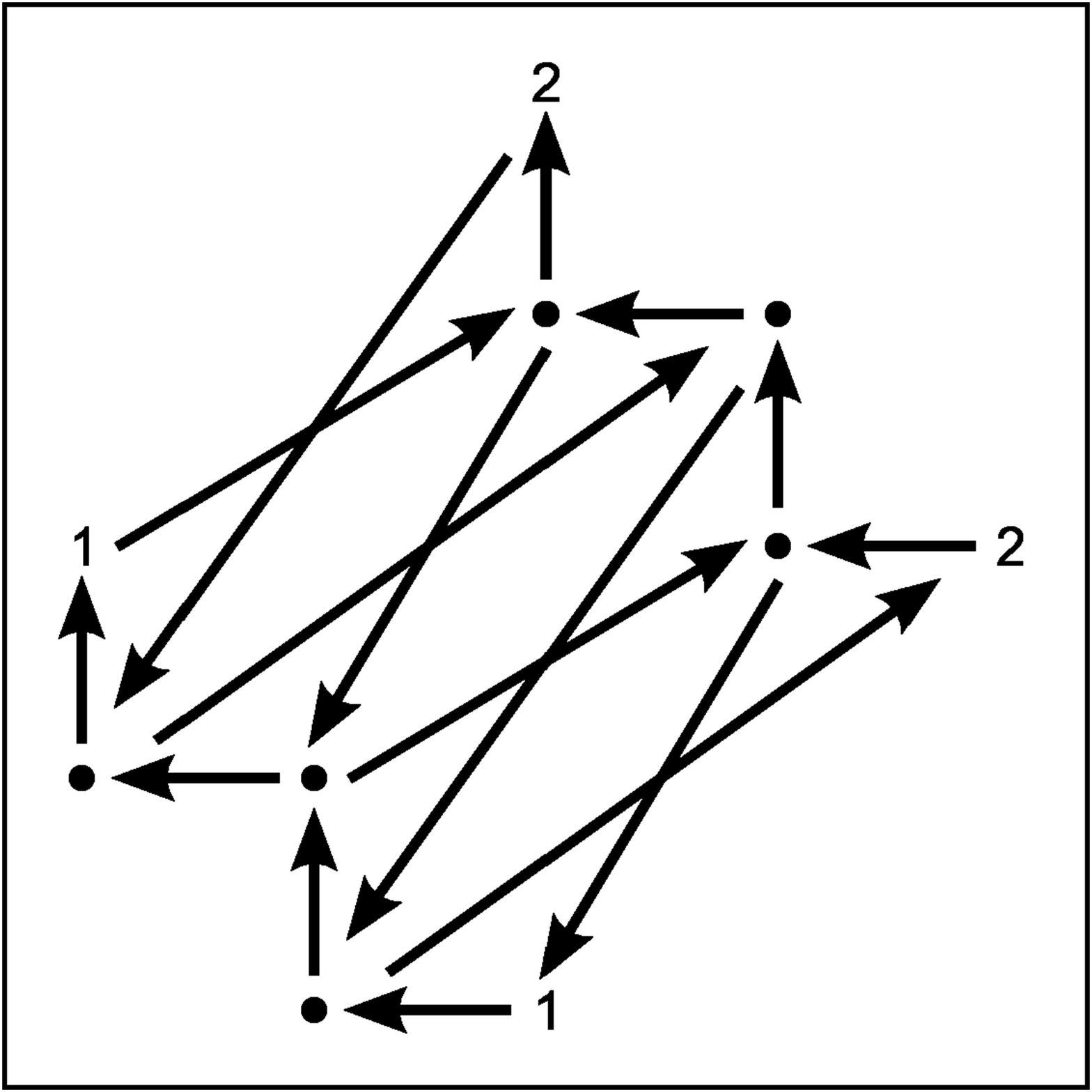}
\end{center}
\caption{Reduction from the dmKdV quiver to the $q$-PVI quiver}
\label{qPVIr}
\end{minipage}
\begin{minipage}{0.5\hsize}
\begin{center}
\includegraphics[width=2cm]{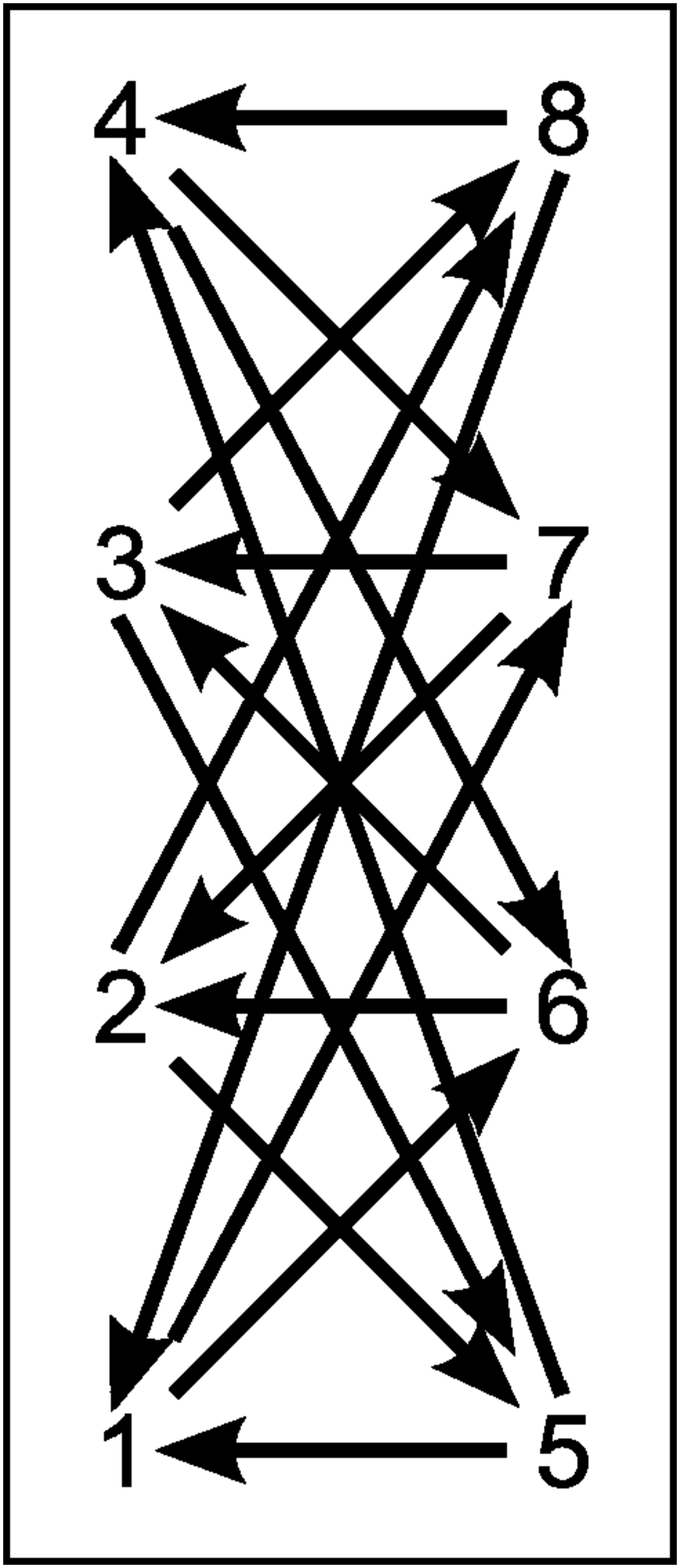}
\end{center}
\caption{The $q$-PVI quiver}
\label{qPVI}
\end{minipage}
\end{figure}

We now construct the quiver of the $q$-Painlev\'e VI equation by a reduction of the dmKdV quiver.
Let $Q_{\rm{PVI}}$ (the $q$-PVI quiver) be the quiver obtained from the $(2,-2)$-reduction of the dmKdV quiver $Q_{\rm{mKdV}}^{1,1}$ (cf. Figures \ref{qPVIr}, \ref{qPVI}).
Let $\nu \in S_8$ be $\nu :(1,2,3,4,5,6,7,8)\mapsto (5,6,7,8,4,3,2,1)$.
$\bm{i}=(1,2,3,4)$ is a $\nu$-period of $Q_{\rm{PVI}}$.
Note that vertices $(1,2,\dots,8)$ correspond to cluster variables $\bm{x}=(w_1,x_1,W_1,X_1,w_2,x_2,W_2,X_2)$ and coefficients $\bm{y}=(y_{1,1},z_{1,1},Y_{1,1},Z_{1,1},y_{2,1},z_{2,1},Y_{2,1},Z_{2,1})$.
We take $(Q_{\rm{PVI}},\bm{x},\bm{y})$ as an initial seed and mutate it in the order $\mu_{\bm{i}},\mu_{\nu(\bm{i})},\mu_{\nu^2(\bm{i})},\dots$.
The new cluster variables are denoted as $w_n\to X_{n+2},\ x_n\to W_{n+2},W_n\to x_{n+2},Z_n\to w_{n+2}$ and the new coefficients are denoted as $\ y_{n,m}\to Z_{n,m+1},\ z_{n,m}\to Y_{n,m+1},\ Y_{n,m}\to z_{n,m+1},\ Z_{n,m}\to y_{n,m+1}$ (in case the coefficient corresponds to the vertex to which the mutation is applied) or $\ y_{n,m}\to y_{n,m+1},\ z_{n,m}\to z_{n,m+1},Y_{n,m}\to Y_{n,m+1},\ Z_{n,m}\to Z_{n,m+1}$ (otherwise).
We put $y_n:=y_{m,n},z_n:=z_{m,n},Y_n:=Y_{m,n},Z_n:=Z_{m,n}\ (n\equiv m\ (\rm{mod}\ 2))$ and obtain the following seeds:
\begin{equation}
\begin{aligned}
\cdots &\overset{\mu_5\mu_6\mu_7\mu_8}{\longleftrightarrow}&
(Q_{\rm{VI}}&;w_1,x_1,W_1,X_1,w_2,x_2,W_2,X_2;\\
&&&\ y_1,z_1,Y_1,Z_1,y_{2,1},z_{2,1},Y_{2,1},Z_{2,1})&\\
&\overset{\mu_4\mu_3\mu_2\mu_1}{\longleftrightarrow}&
(\nu(Q_{\rm{VI}})&;X_3,W_3,x_3,w_3,w_2,x_2,W_2,X_2;\\
&&&\ Z_{1,2},Y_{1,2},z_{1,2},y_{1,2},y_2,z_2,Y_2,Z_2)&\\
&\overset{\mu_8\mu_7\mu_6\mu_5}{\longleftrightarrow}&
(Q_{\rm{VI}}&;X_3,W_3,x_3,w_3,X_4,W_4,x_4,w_4;\\
&&&\ Z_3,Y_3,z_3,y_3,Z_{2,3},Y_{2,3},z_{2,3},y_{2,3})&\\
&\overset{\mu_1\mu_2\mu_3\mu_4}{\longleftrightarrow}&
(\nu(Q_{\rm{VI}})&;w_5,x_5,W_5,X_5,X_4,W_4,x_4,w_4;\\
&&&\ y_{1,4},z_{1,4},Y_{1,4},Z_{1,4},Z_4,Y_4,z_4,y_4)&
\overset{\mu_5\mu_6\mu_7\mu_8}{\longleftrightarrow}\cdots.
\end{aligned}
\end{equation}
We obtain the following proposition from the definition of mutation \eqref{clus}.
\begin{prop}
Consider the coefficient-free cluster algebra $\mathcal{A}(Q_{\rm{PVI}},\bm{x})$.
For any $n\in \mathbb{Z}$, the cluster variables $w_n,x_n,W_n,X_n$ satisfy the bilinear equations:
\begin{equation}\label{eq:qPVIb}
\begin{aligned}
w_{n+2}X_n&=x_{n+1}W_{n+1}+w_{n+1}X_{n+1},\\
x_{n+2}W_n&=w_{n+1}X_{n+1}+x_{n+1}W_{n+1},\\
W_{n+2}x_n&=X_{n+1}w_{n+1}+W_{n+1}x_{n+1},\\
X_{n+2}w_n&=W_{n+1}x_{n+1}+X_{n+1}w_{n+1}.
\end{aligned}
\end{equation}
\end{prop}
The bilinear equations \eqref{eq:qPVIb} are obtained from the discrete mKdV equation \eqref{eq:dmKdV} by imposing the reduction condition $w_{n+2}^{m-2}=w_n^m,x_{n+2}^{m-2}=x_n^m$ and $w_n:=w_n^0,x_n:=x_n^0,W_n:=w_{n-1}^1,X_n:=x_{n-1}^1$.
This reduction of the difference equation corresponds to the $(2,-2)$-reduction of the dmKdV quiver.
Moreover, the coefficients satisfy the $q$-Painlev\'e VI equation.
\begin{thm}
Consider the cluster algebra with coefficient $\mathcal{A}(Q_{\rm{PVI}},\bm{x},\bm{y})$.
For any $n\in \mathbb{Z}$, the coefficients $y_n,z_n,Y_n,Z_n$ satisfy the equations:
\begin{equation}\label{eq:qPVI}
\begin{aligned}
y_{n+1}y_{n-1}&=c_2^{-1}c_3^2c_5^2c_1^{2n}
\frac{(y_n+1)
\left(c_2c_3^{-1}c_4^{(-1)^{n+1}}c_5^{-1}c_6^{(-1)^{n+1}}y_n+1\right)}{\left(y_n+c_3c_4^{(-1)^n}c_1^n\right)
\left(y_n+c_5c_6^{(-1)^n}c_1^n\right)},\\
z_{n+1}z_{n-1}&=c_2c_3^2c_5^{-2}c_1^{2n}
\frac{(z_n+1)
\left(c_3^{-1}c_4^{(-1)^{n+1}}c_5c_6^{(-1)^n}z_n+1\right)}{\left(z_n+c_3c_4^{(-1)^n}c_1^n\right)
\left(z_n+c_2c_5^{-1}c_6^{(-1)^{n+1}}c_1^n\right)},\\
Y_{n+1}Y_{n-1}&=c_2c_3^{-2}c_5^2c_1^{2n}
\frac{(Y_n+1)
\left(c_3c_4^{(-1)^n}c_5^{-1}c_6^{(-1)^{n+1}}Y_n+1\right)}{\left(Y_n+c_2c_3^{-1}c_4^{(-1)^{n+1}}c_1^n\right)
\left(Y_n+c_5c_6^{(-1)^n}c_1^n\right)},\\
Z_{n+1}Z_{n-1}&=c_2^3c_3^{-2}c_5^{-2}c_1^{2n}
\frac{(Z_n+1)\left(c_2^{-1}c_3c_4^{(-1)^n}c_5c_6^{(-1)^n}Z_n+1\right)}{\left(Z_n+c_2c_3^{-1}c_4^{(-1)^{n+1}}c_1^n\right)
\left(Z_n+c_2c_5^{-1}c_6^{(-1)^{n+1}}c_1^n\right)},
\end{aligned}
\end{equation}
where $c_1,c_2,\dots,c_6$ are conserved quantities
\begin{equation}
\begin{aligned}
c_1^2&=\frac{y_{n+1}z_{n+1}Y_{n+1}Z_{n+1}}{y_nz_nY_nZ_n},\quad&
c_2&=y_nz_nY_nZ_nc_1^{-2n},&\\
c_3^2&=y_{2n+1}z_{2n+1}y_{2n}z_{2n}c_1^{-(4n+1)},\quad&
c_4^2&=\frac{y_{2n}z_{2n}}{y_{2n+1}z_{2n+1}}c_1,&\\
c_5^2&=y_{2n+1}Y_{2n+1}y_{2n}Y_{2n}c_1^{-(4n+1)},\quad&
c_6^2&=\frac{y_{2n}Y_{2n}}{y_{2n+1}Y_{2n+1}}c_1.
\end{aligned}
\end{equation}
\end{thm}
If we put $f_n:=y_{2n},g_n:=y_{2n+1}$, we obtain
\begin{equation}
\begin{aligned}
f_{n+1}f_n&=c_2^{-1}c_3^2c_5^2c_1^{4n+2}
\frac{(g_n+1)
\left(c_2c_3^{-1}c_4c_5^{-1}c_6g_n+1\right)}{\left(g_n+c_3c_4^{-1}c_1^{2n+1}\right)
\left(g_n+c_5c_6^{-1}c_1^{2n+1}\right)},\\
g_ng_{n-1}&=c_2^{-1}c_3^2c_5^2c_1^{4n}
\frac{(f_n+1)
\left(c_2c_3^{-1}c_4^{-1}c_5^{-1}c_6^{-1}f_n+1\right)}{\left(f_n+c_3c_4c_1^{2n}\right)
\left(f_n+c_5c_6c_1^{2n}\right)}
\end{aligned}
\end{equation}
from \eqref{eq:qPIII}.
These equations are nothing but the $q$-Painlev\'e VI equation \cite{8}.
\begin{proof}
By the definition of mutation \eqref{coef}, the coefficients $y_{n,m},z_{n,m},Y_{n,m},Z_{n,m}$ satisfy
\begin{equation}
\begin{aligned}
y_n&=y_{n,n-1}(y_{n-1}+1)\left(z_{n-1}^{-1}+1\right)^{-1}
\left(Y_{n-1}^{-1}+1\right)^{-1}(Z_{n-1}+1),\\
y_{n,n-1}&=Z_{n-2}^{-1},\\
z_n&=z_{n,n-1}(z_{n-1}+1)\left(y_{n-1}^{-1}+1\right)^{-1}
\left(Z_{n-1}^{-1}+1\right)^{-1}(Y_{n-1}+1),\\
z_{n,n-1}&=Y_{n-2}^{-1},\\
Y_n&=Y_{n,n-1}(Y_{n-1}+1)\left(Z_{n-1}^{-1}+1\right)^{-1}
\left(y_{n-1}^{-1}+1\right)^{-1}(z_{n-1}+1),\\
Y_{n,n-1}&=z_{n-2}^{-1},\\
Z_n&=Z_{n,n-1}(Z_{n-1}+1)\left(Y_{n-1}^{-1}+1\right)^{-1}
\left(z_{n-1}^{-1}+1\right)^{-1}(y_{n-1}+1),\\
Z_{n,n-1}&=y_{n-2}^{-1},
\end{aligned}
\end{equation}
where we consider the index $n$ of the coefficients $y_{n,m},z_{n,m},Y_{n,m},Z_{n,m}$ as $n\in\mathbb{Z}/2\mathbb{Z}$.
The equations only for $y_n,z_n,Y_n,Z_n$ we obtain, are:
\begin{align}
\label{eq:qPVIy1}
y_{n+2}&=\frac{(y_{n+1}+1)(Z_{n+1}+1)}{\left(z_{n+1}^{-1}+1\right)\left(Y_{n+1}^{-1}+1\right)Z_n},\\
\label{eq:qPVIy2}
z_{n+2}&=\frac{(z_{n+1}+1)(Y_{n+1}+1)}{\left(y_{n+1}^{-1}+1\right)\left(Z_{n+1}^{-1}+1\right)Y_n},\\
\label{eq:qPVIy3}
Y_{n+2}&=\frac{(Y_{n+1}+1)(z_{n+1}+1)}{\left(Z_{n+1}^{-1}+1\right)\left(y_{n+1}^{-1}+1\right)z_n},\\
\label{eq:qPVIy4}
Z_{n+2}&=\frac{(Z_{n+1}+1)(y_{n+1}+1)}{\left(Y_{n+1}^{-1}+1\right)\left(z_{n+1}^{-1}+1\right)y_n}.
\end{align}
We have
\begin{align}
\label{eq:1}
\frac{y_{n+2}z_{n+2}Y_{n+2}Z_{n+2}}{y_{n+1}z_{n+1}Y_{n+1}Z_{n+1}}
&=\frac{y_{n+1}z_{n+1}Y_{n+1}Z_{n+1}}{y_nz_nY_nZ_n},\\
\label{eq:2}
y_{n+2}z_{n+2}Y_nZ_n&=y_{n+1}z_{n+1}Y_{n+1}Z_{n+1},\\
\label{eq:3}
y_{n+2}Y_{n+2}z_nZ_n&=y_{n+1}z_{n+1}Y_{n+1}Z_{n+1}
\end{align}
from \eqref{eq:qPVIy1}$\times$\eqref{eq:qPVIy2}$\times$\eqref{eq:qPVIy3}$\times$\eqref{eq:qPVIy4}, 
\eqref{eq:qPVIy1}$\times$\eqref{eq:qPVIy2}, and \eqref{eq:qPVIy1}$\times$\eqref{eq:qPVIy3} respectively, and we obtain
\begin{equation}
\frac{y_{n+1}z_{n+1}Y_{n+1}Z_{n+1}}{y_nz_nY_nZ_n}=c_1^2
\end{equation}
from \eqref{eq:1}, where $c_1$ is a constant.
In fact, 
\begin{equation}\label{eq:4}
y_nz_nY_nZ_n=c_2c_1^{2n}
\end{equation}
from $y_{n+1}z_{n+1}Y_{n+1}Z_{n+1}=c_1^2y_nz_nY_nZ_n$, where $c_2$ is a constant.
Eliminating $Y_n,Z_n$ from \eqref{eq:2} and \eqref{eq:4} we obtain $y_{n+2}z_{n+2}=c_1^2y_nz_n$.
If we eliminate $z_n,Z_n$ from \eqref{eq:3} and \eqref{eq:4} we obtain $y_{n+2}Y_{n+2}=c_1^2y_nY_n$.
Therefore, 
\begin{equation}\label{eq:5}
y_nz_n=c_3c_4^{(-1)^n}c_1^n,\quad
y_nY_n=c_5c_6^{(-1)^n}c_1^n,
\end{equation}
where $c_3,c_4,c_5,c_6$ are constants.
Eliminating $z_n,Y_n$ from \eqref{eq:4} and \eqref{eq:5} we obtain
\begin{equation}\label{eq:6}
y_n=c_2^{-1}c_3c_4^{(-1)^n}c_5c_6^{(-1)^n}Z_n.
\end{equation}
Finally, eliminating $z_n,Y_n,Z_n$ from \eqref{eq:5}, \eqref{eq:6}, and \eqref{eq:qPVIy1} we obtain the $q$-Painlev\'e VI equation \eqref{eq:qPVI}.
The equations for $z_n,Y_n,Z_n$ are the same as that for $y_n$.
\qed
\end{proof}

\section{Conclusion}

We have shown that cluster variables can satisfy the discrete KdV equation, the Hirota-Miwa equation, the discrete mKdV equation, and the discrete Toda equation, if we take appropriate quivers of initial seeds.
We have also shown that the coefficients of certain cluster algebras satisfy the $q$-Painlev\'e I,II,III,VI equations.
These cluster algebras are obtained from a reduction of the quivers for some integrable partial difference equations.
The $q$-Painlev\'e III equation \eqref{eq:qPIII} and the $q$-Painlev\'e VI equation \eqref{eq:qPVI} are classified as type $(A_2+A_1)^{(1)}$ and $D_5^{(1)}$ in the classification of root systems \cite{8}.
So far we have not obtained the $q$-discrete Painlev\'e equations of type $A_4^{(1)}$ and $E_n^{(1)}$ from cluster algebras.
In the future, we wish to clarify the relations between these equations and cluster algebras.
Quivers of higher order analogue of $q$-Painlev\'e I,II equations are obtained in \cite{9}.
To obtain quivers of higher order analogues of the $q$-Painlev\'e III,VI equations is also a problem we wish to address in the future.

\end{document}